\documentclass[
aps,
prd,
twocolumn,
nofootinbib,
superscriptaddress,
floatfix
]{revtex4-2}
\usepackage{bm}
\usepackage{indentfirst}
\usepackage{amsmath}
\usepackage{graphicx}
\usepackage{float}
\usepackage{amssymb}
\usepackage{psfrag}
\usepackage{hyperref}
\usepackage{tensor}
\usepackage{braket}
\usepackage{multirow}
\usepackage{verbatim}
\usepackage{enumerate}
\usepackage{ulem}
\usepackage{cancel}
\usepackage{booktabs}
\hypersetup{
	colorlinks=true,
	linkcolor=red,
	citecolor=blue,
    urlcolor=blue,
}
\allowdisplaybreaks[1]

 \newcommand{\bq}{\begin{equation}}
 \newcommand{\eq}{\end{equation}}
 \newcommand{\bqn}{\begin{eqnarray}}
 \newcommand{\eqn}{\end{eqnarray}}
 \newcommand{\nb}{\nonumber}

\begin{document}
 \title{Multi-orbit S-star probes of Lorentz violation at the Galactic Center}

\author{Jia-Yuan Fang}
\email{fangjiayuan26@mails.ucas.ac.cn}
\affiliation{School of Fundamental Physics and Mathematical Sciences, Hangzhou Institute for Advanced Study, University of Chinese Academy of Sciences, Hangzhou 310024, China}
\affiliation{School of Physical Sciences, University of Chinese Academy of Sciences, No.19A Yuquan Road, Beijing 100049, China}
\affiliation{Institute of Theoretical Physics, Chinese Academy of Sciences, Beijing 100190, China}

\author{Jian-Ming Yan}
\email{yanjm@ucas.ac.cn}
\thanks{Corresponding author.}
\affiliation{School of Fundamental Physics and Mathematical Sciences, Hangzhou Institute for Advanced Study, University of Chinese Academy of Sciences, Hangzhou 310024, China}

\author{Zong-Kuan Guo}
\email{guozk@itp.ac.cn}
\affiliation{Institute of Theoretical Physics, Chinese Academy of Sciences, Beijing 100190, China}
\affiliation{School of Physical Sciences, University of Chinese Academy of Sciences, No.19A Yuquan Road, Beijing 100049, China}
\affiliation{School of Fundamental Physics and Mathematical Sciences, Hangzhou Institute for Advanced Study, University of Chinese Academy of Sciences, Hangzhou 310024, China}

\begin{abstract}
We investigate the potential of Galactic-center S-star observations to probe Lorentz violation in the minimal pure-gravity Standard-Model Extension (SME), focusing on parameter degeneracies associated with anisotropic orbital signatures. We construct simulated astrometric and spectroscopic observations of S2 and S55 using fiducial orbital and Galactic-center parameters based on the multi-star solution from the GRAVITY Collaboration and evolve the stellar trajectories using the leading Lorentz-violating two-body dynamics supplemented by the Schwarzschild first-post-Newtonian (1PN) correction. Separate Fisher-matrix analyses of S2 and S55 each resolve seven of the eight independent SME directions, while their joint analysis resolves all eight directions by combining complementary projections from the differently oriented orbits. A subsequent Markov chain Monte Carlo (MCMC) analysis confirms the Fisher forecasts and reaches characteristic 95\% credible-level sensitivities of approximately $10^{-3}$-$10^{-2}$ for the time-space components and $10^{-5}$-$10^{-4}$ for the spatial components. Our results demonstrate that combining S-stars with different orbital orientations is effective for breaking geometrical degeneracies and probing the tensor structure of Lorentz violation in the Galactic-center gravitational field.

\end{abstract}

\maketitle
\vspace*{0.5cm}

\section{Introduction}
Lorentz symmetry is a fundamental property of both the Standard Model (SM) of particle physics and general relativity (GR). These two field theories provide profound insights into nature and determine the way we understand the physical world. Despite the remarkable experimental success of both theories, how to unify the SM and GR near the Planck scale remains an open question. Various candidate fundamental theories have therefore motivated the possibility that Lorentz symmetry could be broken at very high energies. Such effects are difficult to probe directly at currently accessible energies; however, suppressed remnants of the symmetry breaking may survive at experimentally accessible energies. A systematic way to search for such residual effects is provided by effective field theory (EFT), which includes an effective Lagrangian constructed from the known low-energy fields and high-energy corrections.

Applying this philosophy to possible Lorentz violation led Colladay and Kostelecký to formulate the Standard-Model Extension (SME), which supplements the conventional SM Lagrangian with Lorentz- and CPT-violating operators while retaining the established field content, internal gauge structure, and observer coordinate covariance~\cite{Colladay:1996iz,Colladay:1998fq}. The SME does not refer to a particular microscopic mechanism; rather, it provides a model-independent framework to describe Lorentz violation. The SME framework was subsequently extended to include gravity by Kostelecký, completing the leading-order SME in Riemann--Cartan spacetime and providing the basis for systematic tests of Lorentz symmetry in gravitational phenomena~\cite{Kostelecky:2003fs,Bailey:2006fd,Hees:2016lyw}.

In the minimal pure-gravity SME, the dominant post-Newtonian Lorentz-violating effects are characterized by the symmetric trace-free coefficients $\bar{s}^{\mu\nu}$, whose nine independent components generate both isotropic and direction-dependent modifications of gravitational dynamics~\cite{Bailey:2006fd}. A broad range of experiments has been used to constrain these coefficients, including lunar laser ranging~\cite{Battat:2007uh,Bourgoin:2016ynf}, atom-interferometric gravimetry~\cite{Muller:2007es}, Gravity Probe B~\cite{Bailey:2013oda}, planetary orbital dynamics~\cite{Hees:2015mga}, Very Long Baseline Interferometry~\cite{LePoncin-Lafitte:2016ocy}, pulsar timing and spin-precession observations~\cite{Shao:2014bfa,Shao:2014oha,Dong:2023nau}, and the Event Horizon Telescope image of Sgr A*~\cite{Khodadi:2022pqh}. More broadly, gravitational-wave observations have recently provided complementary tests of Lorentz violation in the
gravitational SME sector~\cite{Gong2023,Hou2024,Wang2025}. These studies have established stringent bounds on many linear combinations of the gravitational-SME coefficients. An important feature of such tests is that an individual experiment is generally sensitive not to every tensor component separately, but to particular combinations determined by its geometry and motion. Therefore, combining systems with different geometrical configurations is particularly valuable for separating otherwise correlated SME coefficients~\cite{Hees:2015mga,Hees:2016lyw,Dong:2023nau}.

The short-period stars orbiting Sagittarius A* (Sgr A*) provide a complementary laboratory for precision tests of gravity. Long-term astrometric and spectroscopic monitoring of these so-called S-stars demonstrated the presence and properties of the compact object at the Galactic Center and enabled increasingly accurate reconstruction of stellar orbits in its gravitational potential~\cite{Schodel:2002py,Ghez:2008ms,Gillessen:2008qv}. In particular, the highly eccentric and relatively short-period orbit of S2 has become one of the most precise probes of orbital dynamics around a supermassive black hole. Near its 2018 pericenter passage, observations detected the combined gravitational-redshift and transverse-Doppler contribution predicted by GR~\cite{GRAVITY:2018ofz}, followed by the detection of the Schwarzschild precession of its orbit~\cite{GRAVITY:2020gka}. Interferometric astrometry with GRAVITY has subsequently extended high-precision orbital measurements beyond S2 to several stars, notably S29, S38, and S55, and therefore improves the determination of the common Galactic-center potential with joint multi-star analyses~\cite{GRAVITY:2021xju,GRAVITY:2024tth}. These developments make the Galactic Center suitable not only for conventional tests of relativistic orbital dynamics, but also for searches for deviations from GR.

S-star observations have been used to investigate a wide variety of extensions and modifications of GR, such as $f(R)$ and $f(R,\phi)$ extended gravity~\cite{Capozziello:2014rva,Benisty:2023qcv}, scalar–tensor gravity~\cite{Borka-Jovanovic:2019mya}, massive gravity~\cite{Jovanovic:2023tcc}, and other theories~\cite{Hees:2017aal,Yan:2023vdg,Yan:2024bim,DeMartino:2023qkl,Zakharov:2014kka,Tan:2024wuk,Monica:2022qkw,GRAVITY2025FifthForce,YangJiangLin2026,YaoLi2026}. In particular, Lorentz-violating gravity has also begun to be investigated with the orbital motion of S-stars~\cite{qi2026probing,Qi:2026npq}. These studies illustrate the potential of stellar orbital observations near Sgr A* to test gravitational physics on scales intermediate between Solar-System experiments and compact-binary systems.

In existing studies that use S-stars to constrain Lorentz-violating gravity~\cite{Junior:2024ety,qi2026probing,Qi:2026npq}, the metrics considered are all spherically symmetric. For the more general form of the gravitational SME, however, the gravitational potential produced by a central body can be anisotropic, which introduces an additional parameter-identification problem. A single orbit samples only a restricted set of projections on the SME parameter space, and as a result, different SME coefficients can produce strongly correlated orbital signatures. In this paper, we investigate the sensitivity of S-star observations to the coefficients \(\bar s^{\mu\nu}\) of the minimal pure-gravity SME, with particular emphasis on the parameter degeneracies associated with individual orbital geometries. We construct simulated astrometric and spectroscopic observations of S2 and S55 using fiducial orbital and Galactic-center parameters based on the GRAVITY multi-star analysis, and calculate the stellar trajectories using the leading gravitational-SME dynamics together with the standard GR first-post-Newtonian (1PN) correction. We first employ separate S2 and S55 Fisher-matrix analyses to identify the combinations of SME coefficients that can be independently resolved by each star. Then we combine S2 and S55, whose orbital planes have substantially different orientations, to test whether the complementary projections of the anisotropic SME tensor can break these degeneracies. Finally, we perform a full MCMC analysis of the joint simulated data to obtain the nonlinear posterior distributions and verify the Fisher-matrix forecasts. Our main objective is to quantify not only the prospective sensitivity of S-star observations to Lorentz violation, but also the role of orbital diversity in disentangling the components of SME.

The remainder of this paper is organized as follows. Sec.~\ref{sec:SME} introduces the minimal gravitational SME, the Lorentz-violating metric, and the corresponding equations of motion in the central potential. Sec.~\ref{sec:mock} describes the coordinate transformation and the forward model used to calculate the simulated S2 and S55 observations. In Sec.~\ref{sec:fisher}, we present the Fisher-matrix analysis and examine the SME parameter degeneracies for S2 alone, S55 alone, and the joint S2+S55 data sets. Sec.~\ref{sec:mcmc} presents the MCMC analysis and the resulting projected constraints on the SME coefficients. We summarize our conclusions in Sec.~\ref{sec:conclusion}.

\section{Gravitational SME and orbital dynamics}\label{sec:SME}
In this section, we use the action of the minimal SME to derive the SME modification of the weak-field Schwarzschild metric. The metric is obtained using the post-Newtonian method in the weak-field limit. In the Riemann spacetime limit, the action of the minimal SME is constructed with three parts~\cite{Bailey:2006fd}:
\begin{equation}\label{action}
    \mathcal{S} = \mathcal{S}_\text{EH}+\mathcal{S}_m+\mathcal{S}_\text{LV},
\end{equation}
where the first term denotes the Einstein-Hilbert action, 
\begin{equation}
    \mathcal{S}_\text{EH}=\frac{1}{16\pi G}\int d^4x\sqrt{-g}\left(R-2\Lambda\right),
\end{equation}
$\mathcal{S}_m$ denotes the general matter action, and $\mathcal{S}_\text{LV}$ is the Lorentz-violating action, given by
\begin{equation}
    \begin{aligned}
    \mathcal{S}_\text{LV}
    &=\frac{1}{16\pi G}\int d^4x\sqrt{-g}\,
      \bigl(-uR+s^{\mu\nu}R^T_{\mu\nu} \\
    &\qquad\qquad
      +t^{\kappa\lambda\mu\nu}C_{\kappa\lambda\mu\nu}\bigr).
    \end{aligned}
\end{equation}
In this expression, $R^T_{\mu\nu}$ is the trace-free Ricci tensor and $C_{\kappa\lambda\mu\nu}$ is the Weyl conformal tensor. $u, s^{\mu\nu}$ and $t^{\kappa\lambda\mu\nu}$ are real coefficients associated with Lorentz violation. Here, $s^{\mu\nu}$ is symmetric and traceless, whereas $t^{\kappa\lambda\mu\nu}$ has the same index symmetries as the Riemann tensor, and is also traceless.

With the SME action in Eq.~\eqref{action}, we are ready to derive the effective linearized field equations. We begin by expanding the SME modified metric
\begin{equation}
    g_{\mu\nu}=\eta_{\mu\nu}+h_{\mu\nu},
\end{equation}
where $\eta_{\mu\nu}$ is the Minkowski background and $h_{\mu\nu}$ is the linear order metric fluctuation. We assume spontaneous Lorentz breaking in the present treatment, which means the SME coefficients can be decomposed into vacuum expectation values and fluctuations around the vacuum.
\begin{align}
    u&=\bar{u}+\tilde{u},\nb\\
    s^{\mu\nu}&=\bar{s}^{\mu\nu}+\tilde{s}^{\mu\nu},\nb\\
    t^{\kappa\lambda\mu\nu}&=\bar{t}^{\kappa\lambda\mu\nu}+\tilde{t}^{\kappa\lambda\mu\nu}.
\end{align}
Here, $\tilde{u}$, $\tilde{s}^{\mu\nu}$, and $\tilde{t}^{\kappa\lambda\mu\nu}$ are linear fluctuations, and higher-order products are also neglected in the following derivations. After varying the action and linearizing, the effective field equations for $h_{\mu\nu}$ are finally obtained in~\cite{Bailey:2006fd}:
\begin{equation}    
    R_{\mu\nu}=\kappa\left(S_m\right)_{\mu\nu}+\Phi^{\bar{u}}_{\mu\nu}+\Phi^{\bar{s}}_{\mu\nu}+\Phi^{\bar{t}}_{\mu\nu},
\end{equation}
where $(S_m)_{\mu\nu}$ is the trace-reversed matter source, and the $\Phi$-terms contain the SME corrections.
\begin{align}
    \Phi^{\bar{u}}_{\mu\nu}
    &=\bar{u}R_{\mu\nu},\quad \Phi^{\bar{t}}_{\mu\nu}=0,\nb\\
    \Phi^{\bar{s}}_{\mu\nu}
    &=\frac{1}{2}\eta_{\mu\nu}\bar{s}^{\alpha\beta}R_{\alpha\beta}
      -2\bar{s}^{\alpha}_{(\mu}R_{\alpha\nu)}\nb\\
    &\quad+\frac{1}{2}\bar{s}_{\mu\nu}R
      +\bar{s}^{\alpha\beta}R_{\alpha\mu\nu\beta}.
\end{align}
It is worth noting that $\bar{t}$ has no contribution to the linear field equations. The $\bar{u}$-coefficient can be absorbed into a rescaling of G (which is equivalent to setting $\bar{u}=0$), so the phenomenologically interesting leading terms are controlled by $\bar{s}^{\mu\nu}$.

To solve the modified linearized field equations, Bailey and Kostelecký introduce the usual Newtonian and post-Newtonian gravitational potentials, plus tensor potentials needed to capture anisotropic Lorentz violation. With these potentials, they obtain a post-Newtonian metric with Lorentz-violating corrections up to $O(3)$\footnote{Here we use post-Newtonian bookkeeping, take the typical velocity in the system as $v \sim O(1)$, and $v^2\sim GM/r\sim U\sim O(2)$. For simplicity, we follow the style of~\cite{Bailey:2006fd} and omit the raising and lowering of indices with the Minkowski metric $\eta_{\mu\nu}$.}.
\begin{align}\label{metric0}
g_{00} &=-1 + 2U + 3\bar{s}^{00}U + \bar{s}^{jk}U^{jk}- 4\bar{s}^{0j}V^j + O(4),\nb\\
g_{0j} &=-\bar{s}^{0j}U-\bar{s}^{0k}U^{jk}-\frac{7}{2}\left(1+\frac{1}{28}\bar{s}^{00}\right)V^j+\frac{3}{4}\bar{s}^{jk}V^k\nb\\
&\quad-\frac{1}{2}\left(1+\frac{15}{4}\bar{s}^{00}\right)W^j+\frac{5}{4}\bar{s}^{jk}W^k\nb\\
&\quad+\frac{9}{4}\bar{s}^{kl}X^{klj}-\frac{15}{8}\bar{s}^{kl}X^{jkl}-\frac{3}{8}\bar{s}^{kl}Y^{klj},\nb\\
g_{jk} &=\delta^{jk}+(2-\bar{s}^{00})\delta^{jk}U\nb\\
&\quad+\left(\bar{s}^{lm}\delta^{jk}-\bar{s}^{jl}\delta^{mk}-\bar{s}^{kl}\delta^{jm}+2\bar{s}^{00}\delta^{jl}\delta^{km}\right)U^{lm},
\end{align}
where
\begin{align}
    U&=G\int d^3x'\frac{\rho(\vec{x}',t)}{|\vec{x}-\vec{x}'|},\nb\\
    U^{jk}&=G\int d^3x'\frac{\rho(\vec{x}',t)(x^k-x'^k)(x^j-x'^j)}{|\vec{x}-\vec{x}'|^3},
\end{align}
and $V^j,W^j,X^{jkl},Y^{jkl}$ are potentials containing velocity of the source~\cite{Bailey:2006fd}.

For a static point source at the origin, we have
\begin{align}
    U=\frac{GM}{r},\quad U^{jk}=\frac{GM}{r}n^jn^k,\quad n^j&=\frac{x^j}{r}.\nb\\
    V^j=W^j=X^{jkl}=Y^{jkl}=0.
\end{align}
After substituting these equations into the post-Newtonian metric in Eq.~\eqref{metric0}, we finally find the Schwarzschild-like metric under the weak-field approximation, written as
\begin{align}\label{metric1}
ds^2&=-f(\mathbf{x})\,dt^2-2A_i(\mathbf{x})\,dt\,dx^i+\gamma_{ij}(\mathbf{x})\,dx^i dx^j ,\nb\\[1em]
f(\mathbf{x})&=1-\frac{2GM}{r}\left[1+\frac{3}{2}\bar{s}^{00}+\frac{1}{2}\bar{s}^{jk}n^j n^k\right],\nb\\[1em]
A_i(\mathbf{x})&=\frac{GM}{r}\left(\bar{s}^{0i}+\bar{s}^{0k}n^i n^k\right),\nb\\[1em]
\gamma_{ij}(\mathbf{x})
&=\delta^{ij}+\frac{GM}{r}
  \bigl[\left(2-\bar{s}^{00}+\bar{s}^{kl}n^k n^l\right)\delta^{ij}\nb\\
&\quad-\bar{s}^{ik}n^k n^j-\bar{s}^{jk}n^k n^i
  +2\bar{s}^{00}n^i n^j\bigr].
\end{align}

In this work, our concern is a two-body system with a large central mass $M$ and a much lighter test particle 
$m$. For a test particle in the gravitational field described by Eq.~\eqref{metric1}, the leading Newtonian and Lorentz-violating contributions to the orbital acceleration are given by
\begin{align}\label{eom1}
     \frac{d^2 x^j}{dt^2} &= -\Gamma^j_{00} - 2\Gamma^j_{0k} v^k +\cdots\notag\\
     &=\frac{1}{2}  \partial^j h_{00}- (\partial_k h_0^j - \partial^j h_{0k})v^k+\cdots\notag\\
     &=-\frac{GM}{r^3}
\bigg[\left(1+\frac{3}{2}\bar{s}^{00}\right)x^j+\frac{3}{2}\bar{s}^{kl}\hat{x}^k\hat{x}^l x^j\notag\\
&\quad-\bar{s}^{jk}x^k-2\bar{s}^{0k}v^k x^j+2\bar{s}^{0j}v^k x^k\bigg]+\cdots .
\end{align}
Here, the ellipsis includes higher-order corrections from both GR and Lorentz violation.

The Schwarzschild precession of S2 has already been detected by the GRAVITY Collaboration~\cite{GRAVITY:2020gka}, motivating the inclusion of the standard GR 1PN correction in our simulated orbital dynamics. Since existing experimental constraints indicate that the SME coefficients are small, we restrict the Lorentz-violating acceleration to the leading terms displayed in Eq.~\eqref{eom1}. Higher-order SME contributions, including corrections involving additional post-Newtonian factors and terms quadratic in the SME coefficients, are neglected. We therefore supplement the leading Newtonian and SME accelerations with the Schwarzschild 1PN contribution
\begin{equation}
\label{eq:gr1pn}
a_{\mathrm{GR,1PN}}^j=\frac{GM}{r^3}\left[\left(\frac{4GM}{r}-v^kv_k\right)x^j+4(x^kv_k)v^j\right].
\end{equation}
The total acceleration used in the numerical integration is obtained by adding Eq.~\eqref{eq:gr1pn} to the explicit Newtonian and leading SME terms in Eq.~\eqref{eom1}, while neglecting higher-order SME corrections.

At the leading SME order displayed in Eq.~\eqref{eom1}, the term proportional to $\bar{s}^{00}$ has exactly the same radial dependence as the Newtonian acceleration and therefore produces an overall rescaling of the gravitational parameter. The same degeneracy can be seen explicitly in the spatial sector. We decompose the spatial SME tensor into its trace-free and isotropic parts:
\begin{equation}
    \bar{s}^{jk}=\bar{s}^{jk}_\text{TF}+\frac{1}{3}\bar{s}_\text{sp}\delta^{jk},
\end{equation}
where $\bar{s}_\text{sp}=\bar{s}^{XX}+\bar{s}^{YY}+\bar{s}^{ZZ}.$ Substitution into the spatial terms of acceleration gives
\begin{align}
    \frac{3}{2}\bar{s}^{kl}\hat{x}^k\hat{x}^l x^j-\bar{s}^{jk}x^k
    &=\frac{3}{2}\bar{s}^{kl}_\text{TF}\hat{x}^k\hat{x}^l x^j
      -\bar{s}^{jk}_\text{TF}x^k\nb\\
    &\quad+\frac{1}{6}\bar{s}_\text{sp}x^j.
\end{align}
Thus, the spatial trace also contributes only to the isotropic component of acceleration and is therefore degenerate with a rescaling of $GM$ at the order considered here. In the minimal gravitational SME, $\bar{s}^{\mu\nu}$ is four-dimensionally traceless, which means $\bar{s}^{TT}=\bar{s}_\text{sp}$. Consequently, combining the explicit $\bar{s}^{TT}$ contribution with the isotropic part of the acceleration yields
\begin{equation}
    a^j_\text{iso}=-\frac{GM}{r^3}\left(1+\frac{5}{3}\bar{s}^{TT}\right)x^j.
\end{equation}
This result is consistent with the isotropic limit of the minimal SME, in which the Newtonian gravitational coupling can be redefined as $G_\text{eff}=G(1+\frac{5}{3}\bar{s}^{TT})$.

At the leading SME order retained in the present orbital model, the isotropic contribution is degenerate with a rescaling of the gravitational parameter $GM$. Higher-order SME effects could in principle modify this degeneracy, but they are not included in our truncated dynamical model. We therefore restrict the analysis to the anisotropic
SME sector by setting
\begin{equation}
    \bar{s}^{TT}=0.
\end{equation}
The four-dimensional trace condition then reduces to the spatial trace-free condition $\bar{s}^{XX}+\bar{s}^{YY}+\bar{s}^{ZZ}=0$. The resulting analysis therefore targets the anisotropic part of the SME gravitational field rather than attempting to place an independent constraint on the isotropic coefficient $\bar{s}^{TT}$.

\section{Mock data for the S-stars}\label{sec:mock}
In this section, we construct simulated astrometric and spectroscopic data for S2 and S55 over a 20-year observing baseline. The fiducial orbital and Galactic-center parameters are taken from the GRAVITY four-star analysis of S2, S29, S38, and S55~\cite{GRAVITY:2024tth}. We select S2 and S55 because they are the shortest-period members of the four-star sample and provide strong orbital signals over the observing baseline. More importantly, their orbital planes have substantially different orientations, allowing them to probe complementary projections of the anisotropic SME tensor and break parameter degeneracies that cannot be resolved by a single orbit. The fiducial parameter values are also taken from the best-fit solution reported in ~\cite{GRAVITY:2024tth}.

\subsection{Coordinate transformation}
The calculation involves several distinct reference frames. The first is a Sun-centered celestial-equatorial frame (SCCEF), in which the SME parameters are normally defined. The second is the Sgr A*-comoving local frame, the frame that gives the observed positions and radial velocities. The third is the perifocal orbital frame, which is associated with the orbital elements to be constrained.

First, we rotate and boost the SME coefficients from the SCCEF to the local observer frame at Sgr A*. Here, we use the matrix form of SME coefficients $\bar{s}_{0i}$ and $\bar{s}_{jk}$ in the SCCEF:
\begin{align}
    \mathbf{s}^\textsf{T}_\odot=
    \begin{bmatrix}
       \bar{s}^{TX} \\
       \bar{s}^{TY} \\
       \bar{s}^{TZ}
    \end{bmatrix},\quad
    \mathbf{S}_\odot=
    \begin{bmatrix}
       \bar{s}^{XX} & \bar{s}^{XY} & \bar{s}^{XZ}\\
       \bar{s}^{XY} & \bar{s}^{YY} & \bar{s}^{YZ}\\
       \bar{s}^{XZ} & \bar{s}^{YZ} & \bar{s}^{ZZ}
    \end{bmatrix}.
\end{align}
Then, we construct the rotation matrix with the right ascension and declination of Sgr A* (RA: $\alpha_{\rm SgrA*}=266.4168^\circ$, Dec: $\delta_{\rm SgrA*}=-29.0078^\circ$),
\begin{equation}
    \begin{aligned}
    \mathbf{R}
    &=\mathbf{R}^\textsf{T}_x
      \left(\frac{\pi}{2}-\delta_{\rm SgrA*}\right) \\
    &\quad\times\mathbf{R}^\textsf{T}_z
      \left(\frac{\pi}{2}+\alpha_{\rm SgrA*}\right),
    \end{aligned}
\end{equation}
where $\mathbf{R}_x,\mathbf{R}_z$ denote the rotation matrices about the x and z-axes, respectively; the transpose is taken because we work in the passive rotation picture. In addition to the spatial rotation, we consider a first-order Lorentz boost associated with the motion of the local Sgr A* frame relative to the SCCEF. The boost velocity is written as
\begin{align}
    \bm{\beta}=
    \begin{bmatrix}
        \beta_E \\
        \beta_N \\
        \beta_L
    \end{bmatrix}
    \approx
    \begin{bmatrix}
        -4.09\times10^{-4} \\
        -7.31\times10^{-4} \\
        -3.70\times10^{-5}
    \end{bmatrix},
\end{align}
where $\mathbf{v_\text{loc}}=c\bm{\beta}_\text{loc}$. The subscripts $E,N,L$ denote east, north, and line of sight, respectively. These values are taken from the analysis of solar motion by Schönrich et al.~\cite{Sch_nrich_2010}, and the proper motion of Sgr A* measured by VLBA and VERA~\cite{Oyama_2024}. The SME coefficients in the Sgr A*-comoving local frame $\mathbf{s}^\textsf{T}_\text{loc}, \mathbf{S}_\text{loc}$ and those in the SCCEF, $\mathbf{s}^\textsf{T}_\odot, \mathbf{S}_\odot$, are related by
\begin{align}
    \mathbf{s}^\textsf{T}_\text{rot}&=\mathbf{R}\mathbf{s}^\textsf{T}_\odot,\quad
    \mathbf{S}_\text{rot}=\mathbf{R}\mathbf{S}_\odot\mathbf{R}^\textsf{T},\notag\\
    \mathbf{s}^\textsf{T}_\text{loc}&=\mathbf{s}^\textsf{T}_\text{rot}-\mathbf{S}_\text{rot}\bm{\beta} - \bar{s}^{00}_{\rm rot}\bm{\beta},\notag\\
    \mathbf{S}_\text{loc}
    &=\mathbf{S}_\text{rot}-\bm{\beta}\mathbf{s}_\text{rot}
      -\mathbf{s}^\textsf{T}_\text{rot}\bm{\beta}^\textsf{T}.
\end{align}
Here, since we adopt $\bar{s}^{TT}=0$ in the SCCEF, we have $\bar{s}^{00}_{\rm rot}=0$, and the third term in the expression of $\mathbf{s}^\textsf{T}_\text{loc}$ is zero. Note that $\bar{s}^{TT}_\text{loc}$ possibly becomes nonzero through the Lorentz boost.

A Keplerian orbit is specified here by the six parameters $a$, $e$, $T_p$, $i$, $\Omega$, and $\omega$. At a given epoch, the position along the orbit is described by the true anomaly $f$. $a$, $e$, and $f$ are usually defined in the perifocal orbital frame, where the x-axis follows the direction Sgr A* points to the pericenter, the y-axis is in the orbital plane, and the z-axis is perpendicular to the orbital plane. Here, $a$ is the semimajor axis, $e$ is the eccentricity, and $f$ is the true anomaly, $f=\theta-\omega$. With initial values of orbital elements, the initial position and velocity of the S-star in the perifocal orbital frame can be determined, by the following equations.
\begin{align}
    r&=\frac{p}{1+e\cos f},\nb\\
    \frac{dr}{dt}&=\sqrt{\frac{GM}{p}}e\sin f,\nb\\
    \frac{d\theta}{dt}&=\sqrt{\frac{GM}{p^3}}\left(1+e\cos f\right)^2.
\end{align}
Here, $p=a\left(1-e^2\right)$, and we use the polar coordinates $(r,\theta)$, which can be conveniently transformed into $(x,y)$ by $x=r \cos f,\,y=r\sin f$.
In this paper, we calculate the EOM~\eqref{eom1} in the local observer frame, but our goal is to constrain orbital elements of S-stars and other parameters. Therefore, we need to do a coordinate transformation from the perifocal orbital frame to the Sgr A*-comoving local frame. The relation between these two coordinate systems is illustrated in Fig.~\ref{fig:orbit}. The orientation of the orbital plane relative to the plane of the sky is described by three orbital angles: $i, \Omega, \omega.$ With these three angles, we construct the Thiele-Innes constants:

%The perifocal orbital frame is used only to construct the Cartesian initial conditions from the osculating Keplerian elements at the reference epoch.

\begin{figure}[t]
\centering
{\includegraphics[width=\columnwidth]{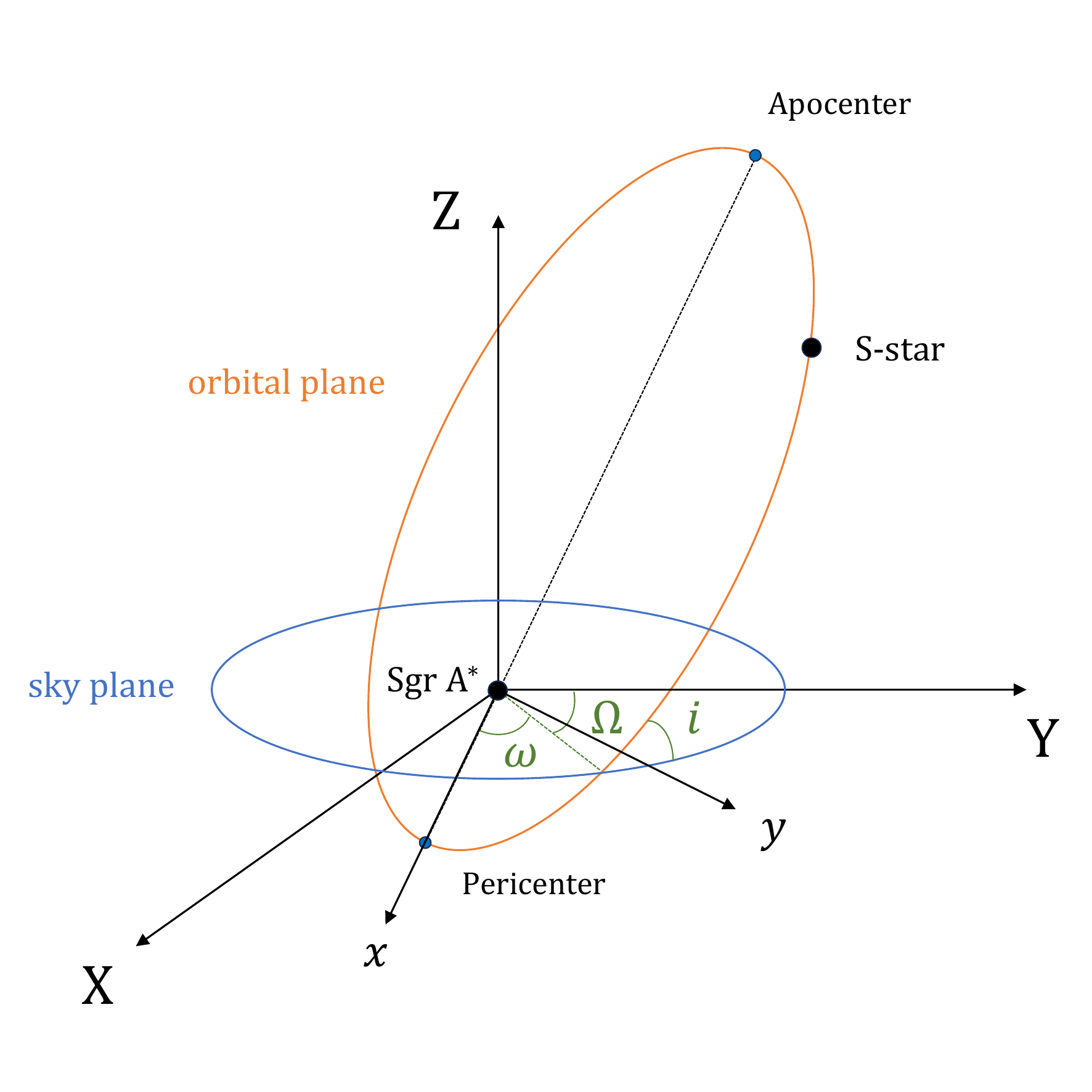}}
\caption{Schematic geometry of an S-star orbit around Sgr A*. 
The blue ellipse represents the sky plane, where X and Y denote the projected right ascension and declination directions, respectively, and Z is the line of sight direction. 
The orange ellipse represents the orbital plane of the S-star. 
The orbital-plane coordinates x and y are defined in the perifocal frame, with the x-axis pointing toward the pericenter and the y-axis lying in the orbital plane in the direction of increasing true anomaly. 
The orientation angles follow the conventional notation: $i$ is the inclination of the elliptical orbit, $\Omega$ is the longitude of the ascending node, and $\omega$ is the argument of pericenter.}\label{fig:orbit}
\end{figure}

\begin{align}
    A&=\cos\Omega\cos\omega-\cos i\sin\Omega\sin\omega,\nb\\
    B&=\sin\Omega\cos\omega+\cos i\cos\Omega\sin\omega,\nb\\
    C&=\sin i\sin\omega,\nb\\
    F&=-\cos\Omega\sin\omega-\cos i\sin\Omega\cos\omega,\nb\\
    G&=-\sin\Omega\sin\omega+\cos i\cos\Omega\cos\omega,\nb\\
    H&=\sin i\cos\omega.
\end{align}
The location of the S-star in the orbital plane $(x,y)$ is then projected onto the sky plane with the coordinates $(X,Y,Z)$,
\begin{align}
    X&=xB+yG,\nb\\
    Y&=xA+yF,\nb\\
    Z&=xC+yH.
\end{align}

\subsection{Mock data sets}
We construct idealized astrometric and spectroscopic data sets for S2 and S55 using the best-fitting parameters obtained from the four-star analysis of S2, S29, S38, and S55 reported by the GRAVITY Collaboration~\cite{GRAVITY:2024tth}. Although the fiducial parameters are obtained from the four-star analysis, only S2 and S55 are simulated because our purpose is to investigate how complementary orbital geometries affect SME parameter identifiability. The parameters include
\begin{equation}\label{parameters}
    \left\{a,e,T_p,i,\Omega,\omega,M,R_0,x_0,y_0,\dot x_0,\dot y_0,v_\text{LSR}\right\},
\end{equation}
where $\{a,e,i,\Omega,\omega\}$ are orbital elements, as defined above, and $T_p$ is the time of pericenter passage. The six parameters of each star are interpreted as osculating Keplerian elements. They are defined at the epochs adopted in the GRAVITY four-star fit,  $t_\text{osc}=2010.35$ for S2 and $t_\text{osc}=2012.0$ for S55. The first six parameters differ between S2 and S55, whereas the remaining parameters are shared by the two stars. $M$ refers to the mass of the central black hole, and $R_0$ denotes the distance between the Earth and the black hole. The positional offsets and drifts $x_0,y_0,\dot x_0,\dot y_0$ refer to epoch 2000.0, and $v_\text{LSR}$ is the constant line-of-sight velocity zero-point of the spectroscopic reference frame. The values and osculating-element epochs follow Table B.1 of the GRAVITY four-star analysis.

We adopt a null Lorentz-violation injection:
\begin{equation}
    \begin{aligned}
    \bar{s}^{TX}=\bar{s}^{TY}=\bar{s}^{TZ}&=0,\\
    \bar{s}^{XY}=\bar{s}^{XZ}=\bar{s}^{YZ}&=0,\\
    \bar{s}^{XX}=\bar{s}^{YY}=\bar{s}^{ZZ}&=0.
    \end{aligned}
\end{equation}
Consequently, any nonzero SME coefficients recovered from an individual noise realization arise from statistical fluctuations or parameter degeneracies rather than from an injected signal. The same fiducial SME coefficients are used for both stars.

The observing cadence is based on the future-observation prescription of Appendix A of~\cite{GRAVITY:2024tth}. Observations are restricted to the approximate Galactic-center observing season between March and September of each year. We assign ten astrometric epochs and three spectroscopic epochs per year. In our implementation, the epochs are uniformly distributed between day 60 and day 273 of the year. The simulated baseline extends from decimal year 2016.53 to 2036.53, immediately following the end of the publicly available data of S2. After applying the limits of this interval to the seasonal sampling, each star has
\begin{equation}
    N_\text{ast}=200,\quad N_\text{RV}=60.
\end{equation}

For each astrometric epoch, independent uncertainties of
\begin{equation}
    \sigma_x=\sigma_y=50\,\mu\text{as}
\end{equation}
are assigned to the two sky coordinates. Each radial-velocity measurement is assigned an uncertainty
\begin{equation}
    \sigma_\text{RV}=10\,\text{km/s}.
\end{equation}
These values correspond to the idealized GRAVITY and ERIS precisions used in the GRAVITY forecast. The noise is assumed to be Gaussian and statistically independent between epochs. The astrometric and spectroscopic data are also independent. The simulations therefore neglect epoch-dependent uncertainties, correlations between right ascension and declination, and instrument-dependent astrometric systematics.

The finite propagation time of light across the stellar orbit is included through the Rømer delay. With the adopted convention that positive $Z$ points away from the observer, the relation between the emission time and the observation time is
\begin{align}
    t_\text{obs}-t_\text{em}=\frac{Z(t_\text{em})}{c},
\end{align}
where $Z(t_\text{em})$ is the distance between the location of each S-star on the orbital plane and the sky plane.

The leading line-of-sight velocity is obtained from the $Z$ component of the Cartesian velocity at the retarded emission time. The model then includes the lowest-order transverse-Doppler and gravitational-redshift terms:
\begin{equation}
    \begin{aligned}
    V_\text{spec}(t_\text{obs})
    &=V_Z(t_\text{em})+\frac{v^2(t_\text{em})}{2c}\\
    &\quad+\frac{GM}{c\,r(t_\text{em})}+v_\text{LSR}.
    \end{aligned}
\end{equation}
Here $v^2/(2c)$ is the special-relativistic transverse-Doppler contribution expressed as an equivalent velocity shift, $GM/(cr)$ is the weak-field gravitational-redshift contribution, and $v_\text{LSR}$ is the common spectroscopic zero point.

\section{Fisher-matrix analysis}\label{sec:fisher}
In this section, we use the Fisher information matrix to examine the local parameter covariance and, in particular, to identify degeneracies among the SME coefficients. Before performing the full MCMC analysis, we compare the local Fisher information obtained from S2 alone, S55 alone, and their joint observations. This comparison allows us to examine not only the parameter degeneracies associated with an individual stellar orbit, but also how the complementary orbital geometries of the two stars help resolve otherwise unconstrained combinations of the SME coefficients. Throughout the Fisher analysis, the observable derivatives
are evaluated using the same orbital model as in Sec.~\ref{sec:mock}, including the Schwarzschild 1PN correction and the leading SME contributions.

For each star, the observable vector is constructed by concatenating the two astrometric coordinates and the spectroscopic radial velocities:
\begin{equation}
    \begin{aligned}
    \bm{\mu(\theta)}
    =\bigl(&x_1,\ldots,x_{N_\text{ast}},\,
             y_1,\ldots,y_{N_\text{ast}},\\
           &v_1,\ldots,v_{N_\text{RV}}\bigr)^\textsf{T}.
    \end{aligned}
\end{equation}
Here, $\bm{\mu(\theta)}$ is evaluated using the forward model described in the previous section, and $\bm{\theta}$ is the vector that includes all orbital and SME parameters. For the joint S2+S55 analysis, the corresponding observable vectors of the two stars are concatenated into a single data vector. Assuming independent Gaussian measurement uncertainties, the likelihood in the vicinity of the fiducial parameter vector $\bm\theta_0$ can be written as
\begin{equation}
    \begin{aligned}
    \ln\mathcal{L}
    &=-\frac{1}{2}\left[\bm{d}-\bm{\mu(\theta)}\right]^\textsf{T}
      \mathbf{C}^{-1}\\
    &\quad\times\left[\bm{d}-\bm{\mu(\theta)}\right]
      +\text{const},
    \end{aligned}
\end{equation}
where $\mathbf{C}$ is diagonal and contains the astrometric and radial-velocity variances specified above. Since these uncertainties are taken to be parameter independent, the Fisher matrix reduces to
\begin{equation}
    \begin{aligned}
    F_{ab}
    &=\frac{\partial\bm{\mu}^\textsf{T}}{\partial\theta_a}
      \mathbf{C}^{-1}
      \frac{\partial\bm{\mu}}{\partial\theta_b}
      \Bigg\rvert_{\bm\theta=\bm\theta_0}.
    \end{aligned}
\end{equation}
Equivalently, defining the whitened Jacobian
\begin{equation}
    A_{\alpha a}=\frac{1}{\sigma_\alpha}\frac{\partial\mu_\alpha}{\partial\theta_a}\Bigg\rvert_{\bm\theta_0},
\end{equation}
we have
\begin{equation}
    \mathbf{F}=\mathbf{A}^\textsf{T}\mathbf{A}.
\end{equation}
For the present parameter-independent Gaussian error model, the Fisher matrix depends on the observing epochs and assumed uncertainties, but not on the particular random-noise realization added to the mock measurements.

The Fisher matrix is evaluated around the null-SME fiducial point:
\begin{align}
    \bar{s}^{\mu\nu}=0.
\end{align}
With the traceless condition on the spatial components, we impose $\bar{s}^{ZZ}=-\bar{s}^{XX}-\bar{s}^{YY}$. The independent SME parameter vector used internally by the Fisher calculation is therefore
\begin{equation}
    \begin{aligned}
    \mathbf{s}=\bigl(&\bar{s}^{TX},\bar{s}^{TY},\bar{s}^{TZ},
                       \bar{s}^{XY},\\
                     &\bar{s}^{XZ},\bar{s}^{YZ},
                       \bar{s}^{XX},\bar{s}^{YY}\bigr).
    \end{aligned}
\end{equation}
Thus, although the figures below display the nine non-TT coefficients, only eight of them are statistically independent in the adopted parameterization.

Since our main interest is in the SME coefficients, we divide the full parameter vector into
\begin{equation}
    \bm\theta=(\bm{\eta},\mathbf{s}),
\end{equation}
where $\mathbf{s}$ contains the eight independent SME coefficients and $\bm{\eta}$ denotes all non-SME parameters. Correspondingly, the whitened Jacobian is written as $(\mathbf{A}_{\eta},\mathbf{A}_{s})$. To isolate the part of the SME signal that cannot be reproduced by variations of the nuisance parameters, we project the SME derivatives onto the subspace orthogonal to the column space of $\mathbf{A}_{\eta}$:
\begin{equation}
    \mathbf{A}_{s,\perp}=\left(\mathbf{I}-\mathbf{P}_{\eta}\right)\mathbf{A}_{s},
\end{equation}
where $\mathbf{P}_{\eta}$ is the orthogonal projector onto the nuisance-parameter subspace. The effective Fisher matrix for the SME sector is then
\begin{equation}
    \mathbf{F}_{s}^\text{eff}=\mathbf{A}_{s,\perp}^\textsf{T}\mathbf{A}_{s,\perp}.
\end{equation}
When the nuisance-parameter Fisher block is nonsingular, this is equivalent to the Schur-complement form
\begin{equation}
    \begin{aligned}
    \mathbf{F}_{s}^\text{eff}
    &=\mathbf{F}_{ss}
      -\mathbf{F}_{s\eta}\mathbf{F}_{\eta\eta}^{-1}
       \mathbf{F}_{\eta s},
    \end{aligned}
\end{equation}
which corresponds to marginalizing over the non-SME parameters in the local Gaussian approximation. The final SVD and noise-aware truncation are then performed only on the projected SME space
\begin{equation}
    \begin{aligned}
    \mathbf{A}_{s,\perp}&=\mathbf{U}\bm\Sigma\mathbf{V}^\textsf{T},\\
    \bm\Sigma&=\operatorname{diag}(s_1,\ldots,s_8),
    \end{aligned}
\end{equation}
where each right singular vector defines a particular linear combination of SME coefficients. Large singular values correspond to combinations that produce distinguishable changes in the observables, whereas very small singular values identify directions that can scarcely be distinguished. Rather than applying an arbitrary relative SVD threshold, our method adopts a noise-aware absolute cutoff:
\begin{equation}
    s_k>s_\text{cut}=\lambda_\text{cut}\delta_B,
\end{equation}
where $\delta_B$ is the numerical perturbation floor obtained from the matrix-level derivative-stability analysis and we take $\lambda_\text{cut}=3$. Only singular modes above this threshold are regarded as numerically resolved. The corresponding rank-limited pseudo-covariance is
\begin{equation}
    \begin{aligned}
    \mathbf{C}_s^+
    &=\mathbf{V}_k
      \operatorname{diag}(s_k^{-2})\Big\rvert_{k\in K}
      \mathbf{V}_k^\textsf{T},
    \end{aligned}
\end{equation}
where $K$ denotes the retained singular modes.

To visualize the covariance structure, we generate Gaussian
samples from the local Fisher approximation and display their
one- and two-dimensional projections using corner plots. For the
full-rank case, the samples are drawn from a multivariate Gaussian
centered at the fiducial point with covariance
$(F_s^{\rm eff})^{-1}$. When rank truncation is required, the corresponding rank-limited pseudo-covariance $C_s^+$ is
used, so that the Gaussian samples lie entirely in the resolved
subspace. These synthetic samples are introduced only to visualize
the Fisher covariance and are not samples from the full posterior
distribution.

This procedure does not change the Fisher information in the nonsingular limit. The advantage of the effective Fisher matrix is that, when rank truncation is necessary, the retained and discarded singular modes are defined directly within the nuisance-marginalized SME subspace. Small singular values can therefore be interpreted specifically as poorly resolved combinations of SME coefficients, rather than as weak directions of the full mixed parameter space. The effective formulation consequently provides a numerically cleaner and more physically transparent basis for diagnosing SME parameter degeneracies and for constructing the rank-limited covariance used below.

\begin{figure*}[!t]
    \centering
    \includegraphics[width=\textwidth]{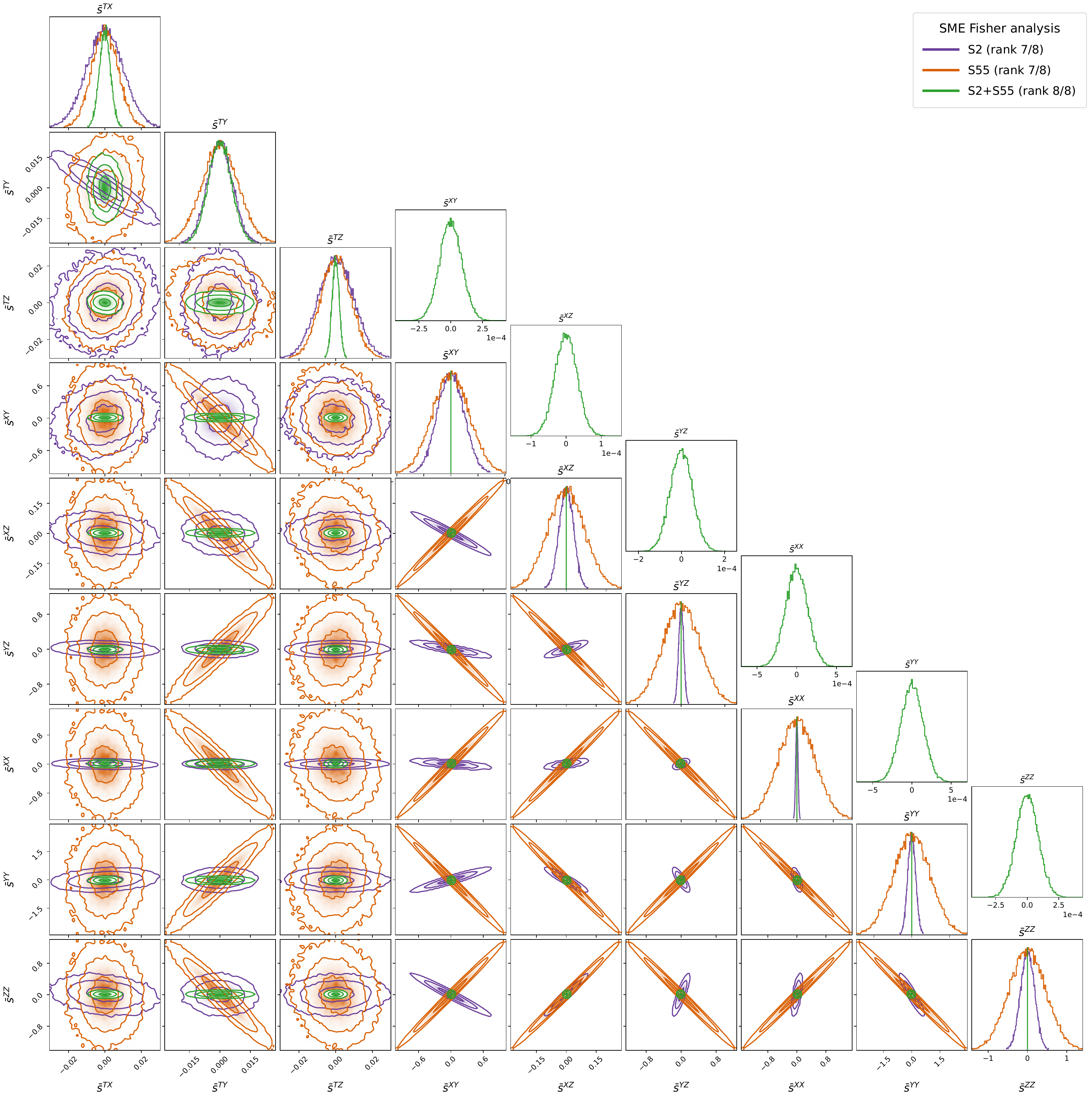}
    \caption{Comparison of the Fisher-matrix covariance structures of the gravitational-SME coefficients obtained from simulated S2-only, S55-only, and joint S2+S55 observations. The purple, orange, and green curves correspond to S2, S55, and the joint analysis, respectively. The main corner panels use common physical parameter ranges for all three cases. The nine displayed coefficients satisfy the spatial trace-free condition, $\bar{s}^{ZZ}=-\bar{s}^{XX}-\bar{s}^{YY}$, leaving eight independent parameters. The corner plots are constructed from synthetic Gaussian samples drawn from the local Fisher approximation. Because the joint constraints on the purely spatial coefficients are much narrower than the corresponding single-star distributions, six additional panels are added to show magnified one-dimensional joint distributions for these coefficients, using rescaled horizontal axes. For the individual stars, only seven singular modes are numerically resolved, and the plotted distributions are constructed from the corresponding rank-limited pseudo-covariances. In contrast, all eight singular modes are resolved in the joint S2+S55 analysis, for which the full Fisher covariance is used. The different orientations of the single-star covariance structures illustrate the complementary parameter information supplied by the two stellar orbits.}\label{fig:fisher_comparison}
\end{figure*}
%The finite widths in these projections do not imply that all eight independent SME directions are constrained. 

\subsection{Single-star constraints}
For S2 alone, seven of the eight independent SME directions are therefore retained, whereas the weakest singular mode lies many orders of magnitude below the numerical resolution threshold. The S2-only effective Fisher matrix consequently has
\begin{equation}
    \text{rank}\left(\mathbf{F}_{s,\text{S2}}^\text{eff}\right)=7\;\text{out of}\;8.
\end{equation}
Note that the unresolved direction is dominated by the time-space coefficients. Thus a single S2 orbit does not contain enough independent geometrical information to determine all SME directions simultaneously. The problem is not simply that each coefficient has a large individual uncertainty; rather, a specific linear combination of the coefficients produces a response that can be almost completely absorbed by changes in the other fitted parameters.

This behavior is physically expected for an anisotropic gravitational perturbation. The SME acceleration contains projections of the background tensor onto the instantaneous orbital geometry, so a single orbital plane samples only a restricted set of tensor projections. Orbital tests of the gravitational SME are therefore naturally sensitive to combinations of coefficients rather than to each tensor component independently.

A similar analysis of S55 alone also yields seven numerically resolved SME singular modes out of eight. Although both individual stellar orbits leave one SME direction unresolved, their covariance structures differ because of their distinct orbital geometries. This is illustrated by the purple and orange contours in Fig.~\ref{fig:fisher_comparison}. For each star alone, several coefficient pairs exhibit highly elongated contours, particularly among the spatial tensor components, demonstrating strong correlations between these coefficients. The sector $\bar{s}^{TX}$-$\bar{s}^{TY}$ also displays a pronounced correlation associated with the weakest SME modes. In particular, the orientations of several strongly correlated parameter combinations differ between the two stars, indicating that they probe complementary projections of the anisotropic SME tensor.

However, it is important to interpret this figure correctly. Because one singular direction is discarded by the noise-aware TSVD, the covariance used to generate each single-star corner plot is a pseudo-covariance in the resolved seven-dimensional subspace. The finite one-dimensional widths shown for the individual coefficients therefore do not imply that all eight independent SME directions are constrained. In the discarded direction, the Fisher uncertainty is formally undefined rather than finite. Similarly, the single-star uncertainties shown in Table~\ref{tab:mcmc_fisher_SME} and Table~\ref{tab:mcmc_fisher_nonSME} are diagnostics and do not represent full marginalized bounds along the unresolved directions. The rank deficiency and the singular-value spectrum, rather than the apparent one-dimensional widths of the corner plot, provide the appropriate diagnostic of the single-star degeneracy.

\subsection{Breaking the degeneracy with S2 and S55}
The situation changes substantially when S2 and S55 are combined. S2 and S55 have markedly different orbital orientations; their orbital angular-momentum directions are separated by approximately $49^\circ$. Consequently, the two stars probe different projections of the same SME tensor. This geometrical complementarity provides the independent directional information needed to disentangle the SME tensor coefficients.

For the joint S2+S55 fit, all eight independent singular modes of SME satisfy $s_k>s_\text{cut}$. Hence,
\begin{equation}
    \text{rank}\left(\mathbf{F}_{s,\text{S2+S55}}^\text{eff}\right)=8,
\end{equation}
and no SME mode must be discarded. Within the local Gaussian approximation, the joint Fisher uncertainties for the independent coefficients are shown in Table~\ref{tab:mcmc_fisher_SME}. In this table, the uncertainty of $\bar{s}^{ZZ}$ is not directly from the effective Fisher matrix. It is constructed from the trace-free condition.

The green contours in Fig.~\ref{fig:fisher_comparison} show the corresponding covariance with the Gaussian approximation. In contrast to the single-star case, all independent SME modes are now retained. On the common parameter scales of the main corner panels, the joint distributions of several spatial coefficients are so narrow that their finite widths are difficult to resolve. The additional zoomed panels therefore show the one-dimensional joint distributions of the spatial SME coefficients on their natural scales. The substantially reduced joint contours in the resolved parameter projections illustrate the improvement obtained by combining the two stellar orbits. Furthermore, the full rank of the joint effective Fisher matrix demonstrates that their complementary information removes the unresolved SME direction present in either single-star analysis. The figure demonstrates that the improvement from the joint S2+S55 result is not merely the statistical gain associated with additional measurements. More importantly, the second orbital plane supplies different directional information about the anisotropic gravitational field, and thereby removes the null combination that is inaccessible to a single S-star.

Taken together, these results motivate the subsequent multi-star MCMC analysis. A single S-star can strongly constrain several combinations of the SME tensor, but it cannot independently determine the complete set of coefficients allowed by our trace-free parameterization. Combining stars on differently oriented orbits probes different projections of the same background tensor and breaks these geometrical degeneracies, allowing all eight independent SME directions to be constrained simultaneously.

\section{Markov Chain Monte Carlo analysis}\label{sec:mcmc}
Following the Fisher-matrix analysis, we perform a full MCMC analysis to sample the joint posterior distribution without relying on the local Gaussian approximation. In this section, we use the Python package $emcee$ to sample the posterior of the eight independent SME coefficients~\cite{Foreman-Mackey:2012any}, together with orbital and Galactic-center parameters in Eq.~\eqref{parameters}. Each single-star Fisher analysis leaves one SME direction almost unconstrained, namely the truncated direction identified by SVD. Therefore, in the subsequent MCMC analysis we present only the results from the joint S2+S55 sampling.

\begin{table*}[htbp]
\centering
\caption{Uniform prior ranges and posterior constraints of the SME coefficients from the joint S2+S55 MCMC analysis, compared with the joint S2+S55 and single-star Fisher-matrix estimates. Here, we sample eight independent coefficients, $\left\{\bar{s}^{TX},\bar{s}^{TY},\bar{s}^{TZ},\bar{s}^{XY},\bar{s}^{XZ},\bar{s}^{YZ},\bar{s}^{XX},\bar{s}^{YY}\right\}$, while the coefficient $\bar{s}^{ZZ}$ is derived from the spatial trace-free condition $\bar{s}^{ZZ}=-\bar{s}^{XX}-\bar{s}^{YY}$. The MCMC intervals denote the 95\% posterior credible intervals, while the joint Fisher results are quoted as symmetric $2\sigma$ uncertainties obtained from the local Gaussian approximation. The single-star entries give $2\sqrt{(C_s^+)_{aa}}$ from the noise-aware TSVD pseudo-covariance in each resolved SME subspace. These quantities have different statistical definitions but are comparable for the nearly Gaussian posteriors considered here.}
\label{tab:mcmc_fisher_SME}

\setlength{\tabcolsep}{3pt}
\renewcommand{\arraystretch}{1.05}

\resizebox{\textwidth}{!}{%
\begin{tabular}{lccccc}
\toprule[0.8pt]
Parameters
& Uniform prior
& \multicolumn{2}{c}{S2+S55 estimate}
& \multicolumn{2}{c}{Single-star estimate} \\
\cmidrule(lr){3-4}\cmidrule(lr){5-6}
&
& MCMC (95\% CI)
& Fisher estimate ($2\sigma$)
& S2 Fisher
& S55 Fisher \\
\midrule[0.5pt]

% SME parameters
$\bar{s}^{TX}$
& $[-3\times10^{-2},\,3\times10^{-2}]$
& $\left(-0.88^{+5.92}_{-5.86}\right)\times10^{-3}$
& $\left(0.00^{+6.22}_{-6.22}\right)\times10^{-3}$
& $\left(0.00^{+2.04}_{-2.04}\right)\times10^{-2}$
& $\left(0.00^{+1.47}_{-1.47}\right)\times10^{-2}$ \\

$\bar{s}^{TY}$
& $[-3\times10^{-2},\,3\times10^{-2}]$
& $\left(-4.03^{+10.72}_{-10.68}\right)\times10^{-3}$
& $\left(0.00^{+11.26}_{-11.26}\right)\times10^{-3}$
& $\left(0.00^{+1.23}_{-1.23}\right)\times10^{-2}$
& $\left(0.00^{+1.82}_{-1.82}\right)\times10^{-2}$ \\

$\bar{s}^{TZ}$
& $[-3\times10^{-2},\,3\times10^{-2}]$
& $\left(2.40^{+3.69}_{-3.69}\right)\times10^{-3}$
& $\left(0.00^{+3.89}_{-3.89}\right)\times10^{-3}$
& $\left(0.00^{+1.94}_{-1.94}\right)\times10^{-2}$
& $\left(0.00^{+1.65}_{-1.65}\right)\times10^{-2}$ \\

$\bar{s}^{XY}$
& $[-10^{-3},\,10^{-3}]$
& $\left(-0.69^{+1.65}_{-1.66}\right)\times10^{-4}$
& $\left(0.00^{+1.73}_{-1.73}\right)\times10^{-4}$
& $\left(0.00^{+4.94}_{-4.94}\right)\times10^{-1}$
& $\left(0.00^{+6.89}_{-6.89}\right)\times10^{-1}$ \\

$\bar{s}^{XZ}$
& $[-10^{-3},\,10^{-3}]$
& $\left(-1.66^{+5.98}_{-5.96}\right)\times10^{-5}$
& $\left(0.00^{+6.29}_{-6.29}\right)\times10^{-5}$
& $\left(0.00^{+7.22}_{-7.22}\right)\times10^{-2}$
& $\left(0.00^{+1.87}_{-1.87}\right)\times10^{-1}$ \\

$\bar{s}^{YZ}$
& $[-10^{-3},\,10^{-3}]$
& $\left(-0.17^{+0.98}_{-0.97}\right)\times10^{-4}$
& $\left(0.00^{+1.03}_{-1.03}\right)\times10^{-4}$
& $\left(0.00^{+1.15}_{-1.15}\right)\times10^{-1}$
& $\left(0.00^{+8.55}_{-8.55}\right)\times10^{-1}$ \\

$\bar{s}^{XX}$
& $[-10^{-3},\,10^{-3}]$
& $\left(-1.29^{+2.66}_{-2.65}\right)\times10^{-4}$
& $\left(0.00^{+2.79}_{-2.79}\right)\times10^{-4}$
& $\left(0.00^{+6.13}_{-6.13}\right)\times10^{-2}$
& $0.00^{+1.03}_{-1.03}$ \\

$\bar{s}^{YY}$
& $[-10^{-3},\,10^{-3}]$
& $\left(0.66^{+2.68}_{-2.68}\right)\times10^{-4}$
& $\left(0.00^{+2.82}_{-2.82}\right)\times10^{-4}$
& $\left(0.00^{+4.16}_{-4.16}\right)\times10^{-1}$
& $0.00^{+1.97}_{-1.97}$ \\

$\bar{s}^{ZZ}$
& \text{--}
& $\left(0.62^{+1.67}_{-1.66}\right)\times10^{-4}$
& $\left(0.00^{+1.76}_{-1.76}\right)\times10^{-4}$
& $\left(0.00^{+3.55}_{-3.55}\right)\times10^{-1}$
& $\left(0.00^{+9.47}_{-9.47}\right)\times10^{-1}$ \\

\bottomrule[0.8pt]
\end{tabular}%
}

\end{table*}

\begin{table*}[t]
\centering
\caption{
Uniform prior ranges and posterior constraints of 19 non-SME parameters from the S2+S55 MCMC
analysis. The MCMC results are compared with the joint S2+S55 and single-star Fisher-matrix estimates, with all Fisher uncertainties quoted at $2\sigma$. The single-star non-SME entries follow the full-parameter raw-Jacobian covariance diagnostics. A dash indicates an orbital parameter of the star not included in that single-star analysis. The non-SME parameters consist of orbital parameters for both stars, $\left\{a_\text{S2},e_\text{S2},T_\text{p,S2},i_\text{S2},\omega_\text{S2},\Omega_\text{S2}\right\}$ and $\left\{a_\text{S55},e_\text{S55},T_\text{p,S55},i_\text{S55},\omega_\text{S55},\Omega_\text{S55}\right\}$; the Galactic-center and reference-frame parameters $\left\{m,d,x_0,y_0,\dot x_0,\dot y_0,v_\text{LSR}\right\}$ are also included.
}
\label{tab:mcmc_fisher_nonSME}

\setlength{\tabcolsep}{3pt}
\renewcommand{\arraystretch}{1.05}

\resizebox{\textwidth}{!}{%
\begin{tabular}{lccccc}
\toprule[0.8pt]
Parameters
& Uniform prior
& \multicolumn{2}{c}{S2+S55 estimate}
& \multicolumn{2}{c}{Single-star estimate} \\
\cmidrule(lr){3-4}\cmidrule(lr){5-6}
&
& MCMC (95\% CI)
& Fisher estimate ($2\sigma$)
& S2 Fisher
& S55 Fisher \\
\midrule[0.5pt]

% S2
$a_{\rm S2}$ (mas)
& $[100,150]$
& $125.013^{+0.012}_{-0.012}$
& $125.020^{+0.013}_{-0.013}$
& $125.020^{+1.396}_{-1.396}$
& \text{--} \\

$e_{\rm S2}$
& $[0.70,0.99]$
& $0.88437^{+0.00014}_{-0.00014}$
& $0.88444^{+0.00015}_{-0.00015}$
& $0.88444^{+0.02162}_{-0.02162}$
& \text{--} \\

$T_{{\rm p},{\rm S2}}$ (yr)
& $[2017.5,2019.0]$
& $2018.3792^{+0.0007}_{-0.0007}$
& $2018.3789^{+0.0007}_{-0.0007}$
& $2018.3789^{+0.8862}_{-0.8862}$
& \text{--} \\

$i_{\rm S2}$ ($^\circ$)
& $[120,150]$
& $134.675^{+0.029}_{-0.028}$
& $134.670^{+0.030}_{-0.030}$
& $134.670^{+0.088}_{-0.088}$
& \text{--} \\

$\omega_{\rm S2}$ ($^\circ$)
& $[55,75]$
& $66.283^{+0.064}_{-0.065}$
& $66.279^{+0.069}_{-0.069}$
& $66.279^{+0.566}_{-0.566}$
& \text{--} \\

$\Omega_{\rm S2}$ ($^\circ$)
& $[200,250]$
& $228.215^{+0.070}_{-0.071}$
& $228.210^{+0.075}_{-0.075}$
& $228.210^{+0.797}_{-0.797}$
& \text{--} \\

\addlinespace

% S55
$a_{\rm S55}$ (mas)
& $[80,130]$
& $104.236^{+0.013}_{-0.013}$
& $104.240^{+0.013}_{-0.013}$
& \text{--}
& $104.240^{+34.341}_{-34.341}$ \\

$e_{\rm S55}$
& $[0.58,0.87]$
& $0.72980^{+0.00021}_{-0.00021}$
& $0.72980^{+0.00023}_{-0.00023}$
& \text{--}
& $0.72980^{+0.08391}_{-0.08391}$ \\

$T_{{\rm p},{\rm S55}}$ (yr)
& $[2008.7,2010.2]$
& $2009.4734^{+0.0007}_{-0.0007}$
& $2009.4738^{+0.0008}_{-0.0008}$
& \text{--}
& $2009.4738^{+0.6944}_{-0.6944}$ \\

$i_{\rm S55}$ ($^\circ$)
& $[145,175]$
& $159.611^{+0.068}_{-0.068}$
& $159.590^{+0.071}_{-0.071}$
& \text{--}
& $159.590^{+0.304}_{-0.304}$ \\

$\omega_{\rm S55}$ ($^\circ$)
& $[318,338]$
& $327.777^{+0.136}_{-0.136}$
& $327.770^{+0.143}_{-0.143}$
& \text{--}
& $327.770^{+22.867}_{-22.867}$ \\

$\Omega_{\rm S55}$ ($^\circ$)
& $[294,344]$
& $319.443^{+0.128}_{-0.130}$
& $319.430^{+0.136}_{-0.136}$
& \text{--}
& $319.430^{+0.529}_{-0.529}$ \\

\addlinespace

% Galactic-center and reference-frame parameters
$m \equiv M/M_\text{fid}$
& $[0.8,1.2]$
& $1.0061^{+0.0055}_{-0.0054}$
& $1.0000^{+0.0058}_{-0.0058}$
& $1.0000^{+0.1839}_{-0.1839}$
& $1.0000^{+0.5634}_{-0.5634}$ \\

$d \equiv R_0/R_{0,\text{fid}}$
& $[0.8,1.2]$
& $1.00211^{+0.00184}_{-0.00181}$
& $1.00000^{+0.00194}_{-0.00194}$
& $1.00000^{+0.00272}_{-0.00272}$
& $1.00000^{+0.01096}_{-0.01096}$ \\

$x_0$ (mas)
& $[-1.0,-0.5]$
& $-0.726^{+0.063}_{-0.062}$
& $-0.724^{+0.066}_{-0.066}$
& $-0.724^{+0.126}_{-0.126}$
& $-0.724^{+0.142}_{-0.142}$ \\

$y_0$ (mas)
& $[-0.5,0]$
& $-0.113^{+0.056}_{-0.055}$
& $-0.119^{+0.059}_{-0.059}$
& $-0.119^{+0.105}_{-0.105}$
& $-0.119^{+0.112}_{-0.112}$ \\

$\dot x_0$ (${\rm km/s}$)
& $[-10,10]$
& $3.099^{+0.087}_{-0.088}$
& $3.070^{+0.091}_{-0.091}$
& $3.070^{+0.158}_{-0.158}$
& $3.070^{+0.181}_{-0.181}$ \\

$\dot y_0$ (${\rm km/s}$)
& $[-10,10]$
& $1.584^{+0.078}_{-0.078}$
& $1.591^{+0.082}_{-0.082}$
& $1.591^{+0.125}_{-0.125}$
& $1.591^{+0.138}_{-0.138}$ \\

$v_\text{LSR}$ (${\rm km/s}$)
& $[-10,10]$
& $-1.60^{+1.74}_{-1.76}$
& $-2.00^{+1.84}_{-1.84}$
& $-2.00^{+3.74}_{-3.74}$
& $-2.00^{+9.01}_{-9.01}$ \\

\bottomrule[0.8pt]
\end{tabular}%
}
\end{table*}

\begin{figure*}[!t]
\centering
\includegraphics[width=\textwidth]{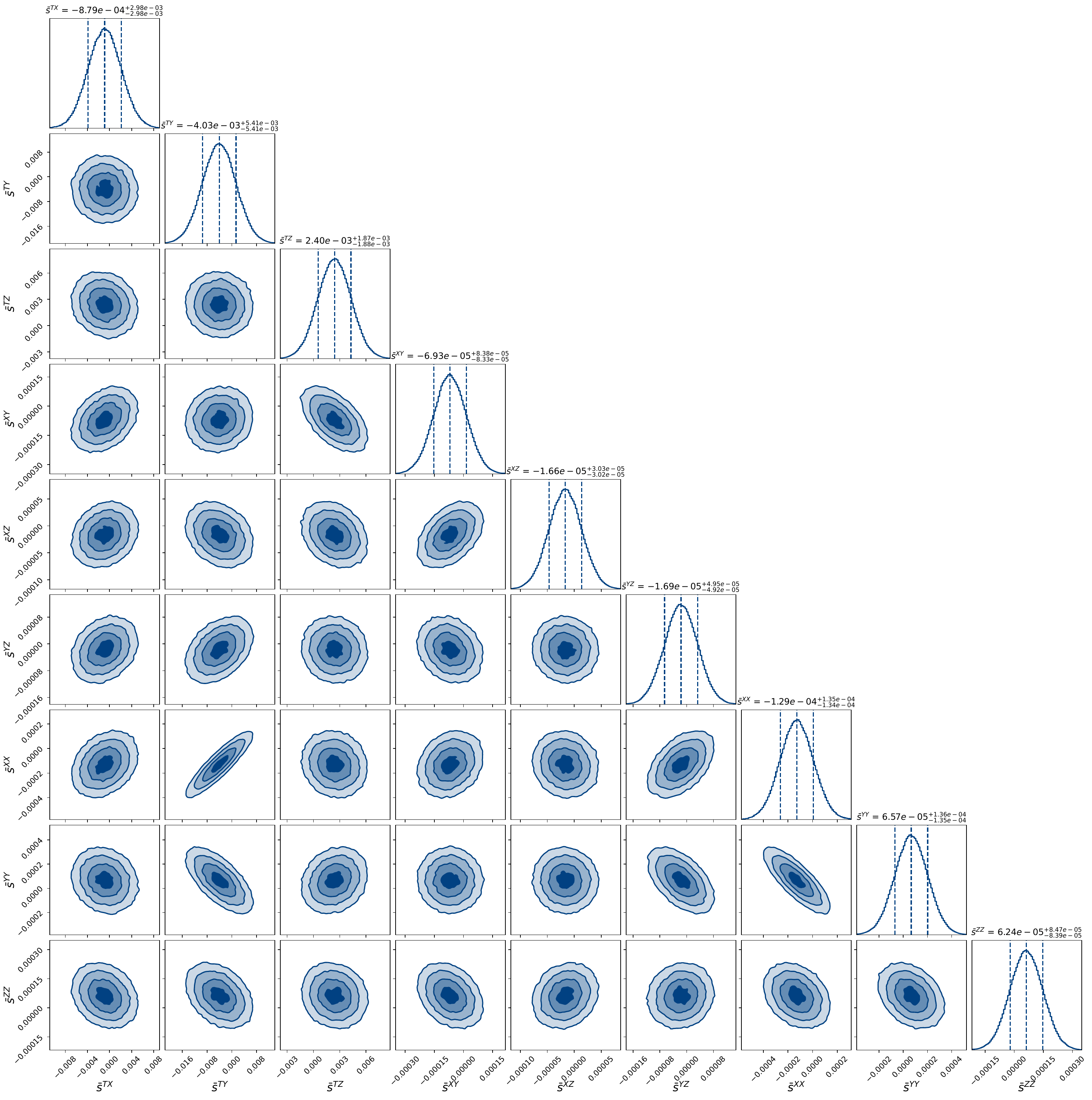}
\caption{Marginalized posterior distributions of the nine displayed SME coefficients obtained from the joint S2+S55 MCMC analysis. Eight coefficients are independently sampled, while $\bar s^{ZZ}$ is reconstructed using the spatial trace-free condition. The one-dimensional distributions show the marginalized posteriors, with vertical dashed lines indicating the 16th, 50th, and 84th percentiles. 
The two-dimensional contours represent the joint posterior correlations among SME coefficients. All coefficients are constrained around the injected zero values within the 95\% credible intervals, while the remaining correlations reflect the geometric complementarity of the two stellar orbits.}\label{fig:MCMC_SME}
\end{figure*}

\subsection{Methods}
The MCMC forward model includes the same Schwarzschild 1PN and leading SME accelerations used to construct the simulated observables. As in the Fisher analysis, we impose $\bar{s}^{ZZ}=-\bar{s}^{XX}-\bar{s}^{YY}$, so that eight independent SME coefficients are directly sampled, while $\bar{s}^{ZZ}$ is reconstructed from the posterior samples.

For a parameter vector $\bm{\theta}$, the posterior is
\begin{align}
    p(\bm{\theta}\mid\bm{d})\propto\mathcal{L}(\bm{d}\mid\bm{\theta})\pi(\bm{\theta}),
\end{align}
where $\pi(\bm{\theta})$ denotes the prior. We adopt bounded uniform priors that encompass the fiducial parameter values. The log-likelihood $\text{log}\mathcal{L}$ consists of two terms:
\begin{align}
    \text{log}\mathcal{L}=\text{log}\mathcal{L}_\text{P}+\text{log}\mathcal{L}_\text{VR}.
\end{align}
Here, the position part $\text{log}\mathcal{L}_\text{P}$ is defined as
\begin{align}
    \log\mathcal{L}_\text{P}
    &=-\frac{1}{2}\sum_i
      \frac{\left(X^i_\text{data}-X^i_\text{theory}\right)^2}
           {\left(\sigma^i_{X,\text{data}}\right)^2}\nb\\
    &\quad-\frac{1}{2}\sum_i
      \frac{\left(Y^i_\text{data}-Y^i_\text{theory}\right)^2}
           {\left(\sigma^i_{Y,\text{data}}\right)^2},
\end{align}
where $X^i_\text{data}$ and $Y^i_\text{data}$ denote the stellar positions measured at epoch $i$, $X^i_\text{theory}$ and $Y^i_\text{theory}$ are the corresponding theoretical positions predicted by the equations of motion, and $\sigma^i_{X,\text{data}}$, $\sigma^i_{Y,\text{data}}$ represent the associated measurement uncertainties. The radial velocity part $\mathcal{L}_\text{VR}$ is given by
\begin{align}
    \text{log}\mathcal{L}_\text{VR}=-\frac{1}{2}\sum_i\frac{\left(V^i_\text{R,data}-V^i_\text{R,theory}\right)^2}{\left(\sigma^i_{V,\text{data}}\right)^2},
\end{align}
in which $V^i_\text{R,data}$ is the observed radial velocity of the star at time $i$, $V^i_\text{R,theory}$ is the corresponding theoretical value derived from our model, and $\sigma^i_{V,\text{data}}$ denotes the uncertainty of this radial velocity measurement.

\subsection{Results}
Following the methods above, we perform the MCMC analysis with mock data of S2 and S55, where all SME coefficients are set to zero. 
The uniform priors of 27 parameters and the corresponding constraints given by the MCMC analysis are shown in Table~\ref{tab:mcmc_fisher_SME} and Table~\ref{tab:mcmc_fisher_nonSME}. Here, the MCMC results are reported as posterior medians within the 95\% credible intervals, while Fisher results are evaluated at the fiducial point with symmetric uncertainties $\pm 2\sigma$. In Table~\ref{tab:mcmc_fisher_nonSME}, the mass and distance are parameterized by the dimensionless scale factors
$m \equiv M/M_\text{fid}$ and $d \equiv R_0/R_{0,\text{fid}}$, respectively, where $M_\text{fid}$ and $R_{0,\text{fid}}$ are fiducial values used in the mock data, $M_\text{fid}=4.2996\times10^6M_\odot,\,R_{0,\text{fid}}=8.2759\text{kpc}$. In Table~\ref{tab:mcmc_fisher_SME}, the coefficient $\bar{s}^{ZZ}$ is not independently sampled and is derived from the spatial trace-free condition $\bar{s}^{ZZ}=-\bar{s}^{XX}-\bar{s}^{YY}$.

The marginalized posterior distributions of the SME coefficients
from the joint S2+S55 MCMC analysis are shown in
Fig.~\ref{fig:MCMC_SME}. The one-dimensional marginalized
posteriors are generally unimodal and approximately Gaussian; the
vertical dashed lines mark the 16th, 50th, and 84th percentiles,
from left to right. The two-dimensional contours reveal several
residual correlations among the SME coefficients, with their
principal directions broadly consistent with the Fisher analysis. Overall, this corner plot demonstrates that combining stars with different orbital geometries substantially reduces parameter degeneracies while allowing the orbital, Galactic-center, and SME parameters to be constrained simultaneously.

The S2+S55 MCMC analysis confirms that the full nonlinear posterior is well constrained and broadly consistent with the Fisher-matrix forecasts. The recovered orbital and Galactic-center parameters remain close to their injected reference values, while the MCMC credible intervals show only modest deviations from the corresponding Fisher $2\sigma$ estimates due to the simulated noise realization and posterior non-Gaussianity. All SME coefficients are consistent with zero within the 95\% credible intervals. The combined orbital information from S2 and S55 constrains all eight independent SME directions, with sensitivities reaching approximately $10^{-3}$ - $10^{-2}$ for the time-space components and $10^{-5}$ - $10^{-4}$ for the spatial components. These results demonstrate the effectiveness of a joint multi-star analysis for testing Lorentz symmetry in the Galactic-center gravitational field.

\section{Conclusion}\label{sec:conclusion}
In this work, we investigated the capability of Galactic-center S-stars to constrain Lorentz violation in the pure-gravity sector of the minimal SME. We constructed a numerical framework that combines the leading Lorentz-violating orbital dynamics with astrometric and spectroscopic observables and applied it to simulated observations of S2 and S55. The fiducial stellar orbits and Galactic-center parameters were based on the GRAVITY multi-star solution, while the mock data were generated assuming a null SME signal. At the order considered here, the isotropic SME contribution is degenerate with a rescaling of the gravitational parameter $GM$. We therefore adopt $\bar{s}^{TT}=0$, which is equivalent to the spatial trace-free condition.

We perform separate Fisher-matrix analyses using S2-only data, S55-only data, and the joint S2+S55 data. After marginalizing over the orbital, Galactic-center, and reference-frame parameters, each single-star effective Fisher matrix resolves only seven of the eight independent SME directions. In each case, one nearly null combination remains, showing that even precise observations of a single orbit cannot independently determine all eight independent SME coefficients considered in this analysis. This behavior arises because a single orbital plane probes only a restricted set of projections of the background tensor. When S2 and S55 are combined, their orbital angular momentum directions allow the two stars to probe different projections of the same anisotropic gravitational field. As a result, all eight independent SME singular modes are resolved in the joint S2+S55 Fisher analysis. This illustrates the central advantage of a multi-star test of Lorentz symmetry at the Galactic Center.

The full MCMC analysis confirms the conclusions obtained from the local Fisher approximation. The orbital parameters of S2 and S55, together with the common Galactic-center and reference-frame parameters, are simultaneously recovered close to their injected values. The marginalized posteriors are approximately Gaussian, and the uncertainties are generally consistent with the Fisher forecasts. Although correlations remain between several SME coefficients, the joint posterior distributions are substantially narrower than the adopted prior ranges once the two stellar orbits are combined. The agreement between the Fisher and MCMC analyses also indicates that, for the simulated data considered here, the local Gaussian approximation provides a useful estimate of the expected parameter sensitivity after the relevant degeneracies have been removed.

Since the mock observations were generated with $\bar{s}^{\mu\nu}=0$, all recovered SME coefficients are consistent with zero within the 95\% credible intervals. The joint S2+S55 analysis reaches characteristic 95\% credible-level sensitivities of approximately $10^{-3}$ - $10^{-2}$ for the time-space coefficients $\bar{s}^{TJ}$ and $10^{-5}$ - $10^{-4}$ for the purely spatial coefficients $\bar{s}^{JK}$. Although these projected constraints do not improve upon the tightest existing limits obtained in the solar system, the S-star orbits can still serve as an independent dataset for studying gravitational Lorentz violation. In this paper, we have considered only S2 and S55 observed for 20 years at the current observational precision; future improvements in astrometric precision and sample size may strengthen these projected constraints.

Overall, our results demonstrate that Galactic-center stellar orbits provide a promising laboratory for probing direction-dependent gravitational effects associated with local Lorentz violation. We find that a single well-measured star leaves an intrinsic SME direction unconstrained, whereas stars moving on differently oriented orbital planes can break this degeneracy. Extending the analysis to a larger sample of accurately monitored S-stars will provide increasingly complete and robust constraints on the SME coefficients with future Galactic-center astrometry and spectroscopy. 

\acknowledgments
This work is supported in part by the National Natural Science Foundation of China Grants No. 12475067 and No. 12235019.  Jian-Ming Yan acknowledges support from the China Postdoctoral Science Foundation Grant No.~2026M793675.

\bibliographystyle{apsrev4-2}
\bibliography{ref}

@article{Bailey:2006fd,
    author = "Bailey, Quentin G. and Kostelecky, V. Alan",
    title = "{Signals for Lorentz violation in post-Newtonian gravity}",
    eprint = "gr-qc/0603030",
    archivePrefix = "arXiv",
    reportNumber = "IUHET-489",
    doi = "10.1103/PhysRevD.74.045001",
    journal = "Phys. Rev. D",
    volume = "74",
    pages = "045001",
    year = "2006"
}

@article{GRAVITY:2024tth,
    author = "Abd El Dayem, Karim and others",
    collaboration = "GRAVITY",
    title = "{Improving constraints on the extended mass distribution in the Galactic center with stellar orbits}",
    eprint = "2409.12261",
    archivePrefix = "arXiv",
    primaryClass = "astro-ph.GA",
    doi = "10.1051/0004-6361/202452274",
    journal = "Astron. Astrophys.",
    volume = "692",
    pages = "A242",
    year = "2024"
}

@article{Oyama_2024,
   title={Trigonometric parallax and proper motion of Sagittarius A* measured by VERA using the new broad-band back-end system OCTAVE-DAS},
   volume={76},
   ISSN={2053-051X},
   DOI={10.1093/pasj/psad088},
   number={2},
    journal={Publ. Astron. Soc. Jpn.},
   publisher={Oxford University Press (OUP)},
   author={Oyama, Tomoaki and Nagayama, Takumi and Yamauchi, Aya and Sakai, Daisuke and Imai, Hiroshi and Honma, Mareki and Asakura, Yu and Hada, Kazuhiro and Hagiwara, Yoshiaki and Hirota, Tomoya and Jike, Takaaki and Kono, Yusuke and Suzuki, Syunsaku and Kobayashi, Hideyuki and Kawaguchi, Noriyuki},
   year={2024},
   month=Jan, pages={163–174} 
}

@article{Sch_nrich_2010,
   title={Local kinematics and the local standard of rest},
   volume={403},
   ISSN={1365-2966},
   DOI={10.1111/j.1365-2966.2010.16253.x},
   number={4},
    journal={Mon. Not. Roy. Astron. Soc.},
   publisher={Oxford University Press (OUP)},
   author={Schönrich, Ralph and Binney, James and Dehnen, Walter},
   year={2010},
   month=Apr, pages={1829–1833} 
}

@article{Bailey:2013oda,
  title = {Limits on Violations of Lorentz Symmetry from Gravity Probe B},
  author = {Bailey, Quentin G. and Everett, Ryan D. and Overduin, James M.},
  year = 2013,
  month = nov,
  journal = {Phys. Rev. D},
  volume = {88},
  eprint = {1309.6399},
  primaryclass = {hep-ph},
  pages = {102001},
  doi = {10.1103/PhysRevD.88.102001},
  archiveprefix = {arXiv}
}

@article{Battat:2007uh,
  title = {Testing for Lorentz Violation: Constraints on Standard-Model Extension Parameters via Lunar Laser Ranging},
  shorttitle = {Testing for Lorentz Violation},
  author = {Battat, James B.R. and Chandler, John F. and Stubbs, Christopher W.},
  year = 2007,
  journal = {Phys. Rev. Lett.},
  volume = {99},
  eprint = {0710.0702},
  primaryclass = {gr-qc},
  pages = {241103},
  doi = {10.1103/PhysRevLett.99.241103},
  archiveprefix = {arXiv}
}

@article{Borka-Jovanovic:2019mya,
  title = {Constraining Scalar-Tensor Gravity Models by S2 Star Orbits around the Galactic Center},
  author = {{Borka-Jovanovi{\'c}}, Vesna and Jovanovi{\'c}, Predrag and Borka, Du{\v s}ko and Capozziello, Salvatore and Gravina, Stefania and D'Addio, Anna},
  year = 2019,
  journal = {Facta Univ. Ser. Phys. Chem. Tech.},
  volume = {17},
  number = {1},
  eprint = {1904.05558},
  primaryclass = {astro-ph.GA},
  pages = {11--20},
  doi = {10.2298/FUPCT1901011B},
  archiveprefix = {arXiv}
}

@article{Bourgoin:2016ynf,
  title = {Testing Lorentz Symmetry with Lunar Laser Ranging},
  author = {Bourgoin, A. and Hees, A. and Bouquillon, S. and {Le Poncin-Lafitte}, C. and Francou, G. and Angonin, M.C.},
  year = 2016,
  month = dec,
  journal = {Phys. Rev. Lett.},
  volume = {117},
  number = {24},
  eprint = {1607.00294},
  primaryclass = {gr-qc},
  pages = {241301},
  doi = {10.1103/PhysRevLett.117.241301},
  archiveprefix = {arXiv}
}

@article{Capozziello:2014rva,
  title = {Constraining Extended Gravity Models by S2 Star Orbits around the Galactic Centre},
  author = {Capozziello, S. and Borka, D. and Jovanovi{\'c}, P. and Jovanovi{\'c}, V. Borka},
  year = 2014,
  month = aug,
  journal = {Phys. Rev. D},
  volume = {90},
  number = {4},
  eprint = {1408.1169},
  primaryclass = {astro-ph.GA},
  pages = {044052},
  doi = {10.1103/PhysRevD.90.044052},
  archiveprefix = {arXiv}
}

@article{Colladay:1996iz,
  title = {CPT Violation and the Standard Model},
  author = {Colladay, Don and Kostelecky, V. Alan},
  year = 1997,
  journal = {Phys. Rev. D},
  volume = {55},
  pages = {6760--6774},
  doi = {10.1103/PhysRevD.55.6760}
}

@article{Colladay:1998fq,
  title = {Lorentz Violating Extension of the Standard Model},
  author = {Colladay, Don and Kostelecky, V. Alan},
  year = 1998,
  journal = {Phys. Rev. D},
  volume = {58},
  pages = {116002},
  doi = {10.1103/PhysRevD.58.116002}
}

@article{DeMartino:2023qkl,
  title = {Constraints on Metric-Affine Gravity Black Holes from the Stellar Motion at the Galactic Center},
  author = {De Martino, Ivan and Della Monica, Riccardo and {Rubiera-Garcia}, Diego},
  year = 2024,
  month = jan,
  journal = {Phys. Rev. D},
  volume = {109},
  number = {2},
  eprint = {2311.08031},
  primaryclass = {gr-qc},
  pages = {024016},
  doi = {10.1103/PhysRevD.109.024016},
  archiveprefix = {arXiv}
}

@article{Dong:2023nau,
  title = {New Limits on the Local Lorentz Invariance Violation of Gravity in the Standard Model Extension with Pulsars},
  author = {Dong, Yiming and Wang, Ziming and Shao, Lijing},
  year = 2024,
  month = apr,
  journal = {Phys. Rev. D},
  volume = {109},
  number = {8},
  eprint = {2311.11038},
  primaryclass = {astro-ph.HE},
  pages = {084024},
  doi = {10.1103/PhysRevD.109.084024},
  archiveprefix = {arXiv}
}

@article{Ghez:2008ms,
  title = {Measuring Distance and Properties of the Milky Way's Central Supermassive Black Hole with Stellar Orbits},
  author = {Ghez, A.M. and Salim, S. and Weinberg, N.N. and others},
  year = 2008,
  journal = {Astrophys. J.},
  volume = {689},
  eprint = {0808.2870},
  primaryclass = {astro-ph},
  pages = {1044--1062},
  doi = {10.1086/592738},
  archiveprefix = {arXiv}
}

@article{Gillessen:2008qv,
  title = {Monitoring Stellar Orbits around the Massive Black Hole in the Galactic Center},
  author = {Gillessen, S. and Eisenhauer, F. and Trippe, S. and Alexander, T. and Genzel, R. and Martins, F. and Ott, T.},
  year = 2009,
  journal = {Astrophys. J.},
  volume = {692},
  eprint = {0810.4674},
  primaryclass = {astro-ph},
  pages = {1075--1109},
  doi = {10.1088/0004-637X/692/2/1075},
  archiveprefix = {arXiv}
}

@article{GRAVITY:2018ofz,
  title = {Detection of the Gravitational Redshift in the Orbit of the Star S2 near the Galactic Centre Massive Black Hole},
   author = {Abuter, R. and Amorim, A. and Anugu, N. and Baub{\"o}ck, M. and Benisty, M. and Berger, J. P. and Blind, N. and Bonnet, H. and Brandner, W. and Buron, A. and others},
   collaboration = {GRAVITY},
  year = 2018,
  journal = {Astron. Astrophys.},
  volume = {615},
  pages = {L15},
  doi = {10.1051/0004-6361/201833718},
  urldate = {2026-08-26},
  langid = {english}
}

@article{GRAVITY:2020gka,
  title = {Detection of the Schwarzschild Precession in the Orbit of the Star S2 near the Galactic Centre Massive Black Hole},
   author = {Abuter, R. and Amorim, A. and Baub{\"o}ck, M. and Berger, J. P. and Bonnet, H. and Brandner, W. and Cardoso, V. and Cl{\'e}net, Y. and {de Zeeuw}, P. T. and Dexter, J. and others},
   collaboration = {GRAVITY},
  year = 2020,
  journal = {Astron. Astrophys.},
  volume = {636},
  pages = {L5},
  doi = {10.1051/0004-6361/202037813},
  urldate = {2026-08-26},
  langid = {english}
}

@article{GRAVITY:2021xju,
  title = {Mass Distribution in the Galactic Center Based on Interferometric Astrometry of Multiple Stellar Orbits},
   author = {Abuter, R. and Aimar, N. and Amorim, A. and Ball, J. and Baub{\"o}ck, M. and Berger, J. P. and Bonnet, H. and Bourdarot, G. and Brandner, W. and Cardoso, V. and others},
   collaboration = {GRAVITY},
  year = 2022,
  journal = {Astron. Astrophys.},
  volume = {657},
  pages = {L12},
  doi = {10.1051/0004-6361/202142465},
  urldate = {2026-08-26},
  langid = {english}
}

@article{Hees:2015mga,
  title = {Testing Lorentz Symmetry with Planetary Orbital Dynamics},
  author = {Hees, Aur{\'e}lien and Bailey, Quentin G. and {Le Poncin-Lafitte}, Christophe and Bourgoin, Adrien and Rivoldini, Attilio and Lamine, Brahim and Meynadier, Fr{\'e}d{\'e}ric and Guerlin, Christine and Wolf, Peter},
  year = 2015,
  month = sep,
  journal = {Phys. Rev. D},
  volume = {92},
  number = {6},
  eprint = {1508.03478},
  primaryclass = {gr-qc},
  pages = {064049},
  doi = {10.1103/PhysRevD.92.064049},
  archiveprefix = {arXiv}
}

@article{Hees:2016lyw,
  title = {Tests of Lorentz Symmetry in the Gravitational Sector},
  author = {Hees, Aur{\'e}lien and Bailey, Quentin G. and Bourgoin, Adrien and Bars, H{\'e}l{\`e}ne Pihan-Le and Guerlin, Christine and {Le Poncin-Lafitte}, Christophe},
  year = 2016,
  journal = {Universe},
  volume = {2},
  number = {4},
  pages = {30},
  doi = {10.3390/universe2040030}
}

@article{Hees:2017aal,
  title = {Testing General Relativity with Stellar Orbits around the Supermassive Black Hole in Our Galactic Center},
  author = {Hees, A. and Do, T. and Ghez, A. M. and Martinez, G. D. and Naoz, S. and Becklin, E. E. and Boehle, A. and Chappell, S. and Chu, D. and Dehghanfar, A. and Kosmo, K. and Lu, J. R. and Matthews, K. and Morris, M. R. and Sakai, S. and Sch{\"o}del, R. and Witzel, G.},
  year = 2017,
  month = may,
   journal = {Phys. Rev. Lett.},
  volume = {118},
  number = {21},
  eprint = {1705.07902},
  primaryclass = {astro-ph},
  pages = {211101},
  issn = {0031-9007, 1079-7114},
  doi = {10.1103/PhysRevLett.118.211101},
  urldate = {2026-03-16},
  archiveprefix = {arXiv}
}

@article{Khodadi:2022pqh,
  title = {Probing the Lorentz Symmetry Violation Using the First Image of Sagittarius A*: Constraints on Standard-Model Extension Coefficients},
  shorttitle = {Probing the Lorentz Symmetry Violation Using the First Image of Sagittarius A*},
  author = {Khodadi, Mohsen and Lambiase, Gaetano},
  year = 2022,
  month = nov,
   journal = {Phys. Rev. D},
  volume = {106},
  number = {10},
  eprint = {2206.08601},
  primaryclass = {gr-qc},
  pages = {104050},
  issn = {2470-0010, 2470-0029},
  doi = {10.1103/PhysRevD.106.104050},
  urldate = {2026-03-16},
  archiveprefix = {arXiv}
}

@article{Kostelecky:2003fs,
  title = {Gravity, Lorentz Violation, and the Standard Model},
  author = {Kostelecky, V. Alan},
  year = 2004,
  journal = {Phys. Rev. D},
  volume = {69},
  pages = {105009},
  doi = {10.1103/PhysRevD.69.105009}
}

@article{LePoncin-Lafitte:2016ocy,
  title = {Lorentz Symmetry and Very Long Baseline Interferometry},
  author = {{Le Poncin-Lafitte}, C. and Hees, A. and Lambert, S.},
  year = 2016,
  month = dec,
  journal = {Phys. Rev. D},
  volume = {94},
  number = {12},
  eprint = {1604.01663},
  primaryclass = {gr-qc},
  pages = {125030},
  doi = {10.1103/PhysRevD.94.125030},
  archiveprefix = {arXiv}
}

@article{Monica:2022qkw,
  title = {Constraining MOdified Gravity with the S2 Star},
  author = {Monica, Riccardo Della and {de Martino}, Ivan and {de Laurentis}, Mariafelicia},
  year = 2022,
  month = feb,
  journal = {Universe},
  volume = {8},
  number = {2},
  eprint = {2206.12699},
  primaryclass = {gr-qc},
  pages = {137},
  doi = {10.3390/universe8020137},
  archiveprefix = {arXiv}
}

@article{Muller:2007es,
  title = {Atom Interferometry Tests of the Isotropy of Post-Newtonian Gravity},
  author = {Muller, Holger and Chiow, Sheng-wey and Herrmann, Sven and Chu, Steven and Chung, Keng-Yeow},
  year = 2008,
  journal = {Phys. Rev. Lett.},
  volume = {100},
  eprint = {0710.3768},
  primaryclass = {gr-qc},
  pages = {031101},
  doi = {10.1103/PhysRevLett.100.031101},
  archiveprefix = {arXiv}
}

@article{Qi:2026npq,
  title = {Detecting Lorentz-Violation Induced by a Tensor Field with S-Star's Motion around Sgr A*},
  author = {Qi, Qi and Sang, Yu and Kuang, Xiao-Mei},
  year = 2026,
  month = jun,
  journal = {Phys. Lett. B},
  volume = {879},
  eprint = {2608.16920},
  primaryclass = {gr-qc},
  pages = {140602},
  doi = {10.1016/j.physletb.2026.140602},
  archiveprefix = {arXiv}
}

@article{qi2026probing,
  author = {Qi, Qi and Sang, Yu and Kuang, Xiao-Mei},
  title = {Probing a Lorentz-violating parameter from orbitalprecession of the S2 star around the galactic centre supermassive black hole},
  journal = {Sci. China Phys. Mech. Astron.},
  volume = {69},
  pages = {260414},
  year = {2026},
  doi = {10.1007/s11433-026-2960-1},
  eprint = {2601.06491},
  archivePrefix = {arXiv},
  primaryClass = {gr-qc}
}

@article{Schodel:2002py,
  title = {A Star in a 15.2 Year Orbit around the Supermassive Black Hole at the Center of the Milky Way},
   author = {Sch{\"o}del, R. and Ott, Thomas and Genzel, R. and others},
  year = 2002,
  journal = {Nature},
  volume = {419},
  eprint = {astro-ph/0210426},
  pages = {694--696},
  doi = {10.1038/nature01121},
  archiveprefix = {arXiv}
}

@article{Shao:2014bfa,
  title = {New Pulsar Limit on Local Lorentz Invariance Violation of Gravity in the Standard-Model Extension},
  author = {Shao, Lijing},
  year = 2014,
  month = dec,
  journal = {Phys. Rev. D},
  volume = {90},
  number = {12},
  eprint = {1412.2320},
  primaryclass = {gr-qc},
  pages = {122009},
  doi = {10.1103/PhysRevD.90.122009},
  archiveprefix = {arXiv}
}

@article{Shao:2014oha,
  title = {Tests of Local Lorentz Invariance Violation of Gravity in the Standard Model Extension with Pulsars},
  author = {Shao, Lijing},
  year = 2014,
  month = mar,
  journal = {Phys. Rev. Lett.},
  volume = {112},
  eprint = {1402.6452},
  primaryclass = {gr-qc},
  pages = {111103},
  doi = {10.1103/PhysRevLett.112.111103},
  archiveprefix = {arXiv}
}

@article{Tan:2024wuk,
  title = {Constraining the Yukawa Gravity with the Post-Newtonian Approximation Using S-Star Orbits around the Supermassive Black Hole in Our Galactic Center},
  author = {Tan, Yuan and Lu, Youjun},
  year = 2024,
  month = feb,
  journal = {Phys. Rev. D},
  volume = {109},
  number = {4},
  eprint = {2402.00333},
  primaryclass = {gr-qc},
  pages = {044047},
  doi = {10.1103/PhysRevD.109.044047},
  archiveprefix = {arXiv}
}

@article{Yan:2023vdg,
  title = {Observational Tests of a Quantum Extension of Schwarzschild Spacetime in Loop Quantum Gravity with Stars in the Galactic Center},
  author = {Yan, Jian-Ming and Liu, Cheng and Zhu, Tao and Wu, Qiang and Wang, Anzhong},
  year = 2023,
  month = apr,
  journal = {Phys. Rev. D},
  volume = {107},
  number = {8},
  eprint = {2302.10482},
  primaryclass = {gr-qc},
  pages = {084043},
  doi = {10.1103/PhysRevD.107.084043},
  archiveprefix = {arXiv}
}

@article{Yan:2024bim,
  title = {Observational Test of {$R^2$} Spacetimes with the S2 Star in the Milky Way Galactic Center},
  author = {Yan, Jian-Ming and Zhu, Tao and {Azreg-A{\"i}nou}, Mustapha and Jamil, Mubasher and Nguyen, Hoang Ky},
  year = 2024,
  journal = {JCAP},
  volume = {2024},
  number = {07},
  eprint = {2405.10559},
  primaryclass = {gr-qc},
  pages = {071},
  doi = {10.1088/1475-7516/2024/07/071},
  archiveprefix = {arXiv}
}

@article{Benisty:2023qcv,
    author = "Benisty, D. and Mifsud, J. and Levi Said, J. and Staicova, D.",
    title = "{Strengthening extended gravity constraints with combined systems: f(R) bounds from cosmology and the galactic center}",
    eprint = "2303.15040",
    archivePrefix = "arXiv",
    primaryClass = "astro-ph.CO",
    doi = "10.1016/j.dark.2023.101344",
    journal = "Phys. Dark Univ.",
    volume = "42",
    pages = "101344",
    year = "2023"
}

@article{Zakharov:2014kka,
    author = "Zakharov, A. F. and Borka, Dusko and Borka Jovanovi{\'c}, Vesna and Jovanovi{\'c}, Predrag",
    title = "{Constraints on $R^n$ gravity from precession of orbits of S2-like stars: a case of a bulk distribution of mass}",
    eprint = "1407.0366",
    archivePrefix = "arXiv",
    primaryClass = "astro-ph.GA",
    doi = "10.1016/j.asr.2014.05.027",
    journal = "Adv. Space Res.",
    volume = "54",
    pages = "1108--1112",
    year = "2014"
}

@article{Jovanovic:2023tcc,
    author = "Jovanovi{\'c}, Predrag and Jovanovi{\'c}, Vesna Borka and Borka, Du{\v{s}}ko and Zakharov, Alexander F.",
    title = "{Improvement of graviton mass constraints using GRAVITY{\textquoteright}s detection of Schwarzschild precession in the orbit of S2 star around the Galactic Center}",
    eprint = "2305.13448",
    archivePrefix = "arXiv",
    primaryClass = "astro-ph.GA",
    doi = "10.1103/PhysRevD.109.064046",
    journal = "Phys. Rev. D",
    volume = "109",
    number = "6",
    pages = "064046",
    year = "2024"
}

@article{Junior:2024ety,
    author = "Junior, Ednaldo L. B. and Junior, Jos{\'e} Tarciso S. S. and Lobo, Francisco S. N. and Rodrigues, Manuel E. and Rubiera-Garcia, Diego and da Silva, Lu{\'\i}s F. Dias and Vieira, Henrique A.",
    title = "{Spontaneous Lorentz symmetry-breaking constraints in Kalb{\textendash}Ramond gravity}",
    eprint = "2405.03291",
    archivePrefix = "arXiv",
    primaryClass = "gr-qc",
    doi = "10.1140/epjc/s10052-024-13619-3",
    journal = "Eur. Phys. J. C",
    volume = "84",
    number = "12",
    pages = "1257",
    year = "2024"
}

@article{Gong2023,
  author = {Gong, Cheng and Zhu, Tao and Niu, Rui and Wu, Qiang
            and Cui, Jing-Lei and Zhang, Xin and Zhao, Wen
            and Wang, Anzhong},
  title = {Gravitational wave constraints on nonbirefringent
           dispersions of gravitational waves due to Lorentz
           violations with GWTC-3 events},
  journal = {Phys. Rev. D},
  volume = {107},
  pages = {124015},
  year = {2023},
  doi = {10.1103/PhysRevD.107.124015},
  eprint = {2302.05077},
  archivePrefix = {arXiv},
  primaryClass = {gr-qc}
}

@article{Hou2024,
  author = {Hou, Shaoqi and Fan, Xi-Long and Zhu, Tao
            and Zhu, Zong-Hong},
  title = {Nontensorial gravitational wave polarizations from
           the tensorial degrees of freedom: Linearized
           Lorentz-violating theory of gravity},
  journal = {Phys. Rev. D},
  volume = {109},
  pages = {084011},
  year = {2024},
  doi = {10.1103/PhysRevD.109.084011},
  eprint = {2401.03474},
  archivePrefix = {arXiv},
  primaryClass = {gr-qc}
}

@article{Wang2025,
  author = {Wang, Qiang and Yan, Jian-Ming and Zhu, Tao
            and Zhao, Wen},
  title = {Modified gravitational wave propagations in linearized
           gravity with Lorentz and diffeomorphism violations
           and their gravitational wave constraints},
  journal = {Phys. Rev. D},
  volume = {111},
  pages = {084064},
  year = {2025},
  doi = {10.1103/PhysRevD.111.084064},
  eprint = {2501.11956},
  archivePrefix = {arXiv},
  primaryClass = {gr-qc}
}

@article{GRAVITY2025FifthForce,
  author = {{GRAVITY Collaboration}},
  title = {Exploring the presence of a fifth force at the
           Galactic Center},
  journal = {Astron. Astrophys.},
  volume = {698},
  pages = {L15},
  year = {2025},
  doi = {10.1051/0004-6361/202554676},
  eprint = {2504.02908},
  archivePrefix = {arXiv},
  primaryClass = {astro-ph.GA}
}

@article{YaoLi2026,
  author = {Yao, Jin-Tao and Li, Xin},
  title = {Probing covariant quantum black holes in effective
           quantum gravity through the orbital motion of the
           S2 star around Sagittarius A*},
  journal = {Phys. Rev. D},
  volume = {113},
  pages = {084003},
  year = {2026},
  doi = {10.1103/688k-qg81}
}

@article{YangJiangLin2026,
  author = {Yang, Bo and Jiang, Chunhua and Lin, Wenbin},
  title = {Probing the Schwarzschild black hole in modified gravity
           by precessing motion of S2 star and OJ 287},
  journal = {Eur. Phys. J. C},
  volume = {86},
  pages = {725},
  year = {2026},
  doi = {10.1140/epjc/s10052-026-15983-8}
}

@article{Foreman-Mackey:2012any,
    author = "Foreman-Mackey, Daniel and Hogg, David W. and Lang, Dustin and Goodman, Jonathan",
    title = "{emcee: The MCMC Hammer}",
    eprint = "1202.3665",
    archivePrefix = "arXiv",
    primaryClass = "astro-ph.IM",
    doi = "10.1086/670067",
    journal = "Publ. Astron. Soc. Pac.",
    volume = "125",
    pages = "306--312",
    year = "2013"
}

\clearpage

\end{document}